\documentclass[12pt]{revtex4}
\usepackage{graphicx}
\usepackage{float}
\usepackage{amsmath}
\usepackage{amsfonts}
\usepackage{wasysym}
\usepackage{tabularx}
\usepackage{color}

\newcommand{\omegabold}{\boldsymbol\omega}

\begin{document}

\title{Development of high vorticity structures in incompressible 3D Euler equations}
\author{D.S. Agafontsev$^{1,2}$, E.A. Kuznetsov$^{2,3,4}$ and A.A. Mailybaev$^{5,6}$}

\affiliation{\small \textit{ $^{1}$ P. P. Shirshov Institute of Oceanology, 36 Nakhimovsky prosp., 117218 Moscow, Russia\\
$^{2}$ Novosibirsk State University, 2 Pirogova, 630090 Novosibirsk, Russia\\
$^{3}$ P.N. Lebedev Physical Institute, 53 Leninsky Ave., 119991 Moscow, Russia\\
$^{4}$ L.D. Landau Institute for Theoretical Physics, 2 Kosygin str., 119334 Moscow, Russia\\
$^{5}$ Instituto Nacional de Matem\'atica Pura e Aplicada -- IMPA, Rio de Janeiro, Brazil\\
$^{6}$ Institute of Mechanics, Lomonosov Moscow State University, Russia}}

\begin{abstract}
We perform the systematic numerical study of high vorticity structures that develop in the 3D incompressible Euler equations from generic large-scale initial conditions. We observe that a multitude of high vorticity structures appear in the form of thin vorticity sheets (pancakes). Our analysis reveals the self-similarity of the pancakes evolution, which is governed by two different exponents $e^{-t/T_{\ell}}$ and $e^{t/T_{\omega}}$ describing compression in the transverse direction and the vorticity growth respectively, with the universal ratio $T_{\ell}/T_{\omega} \approx 2/3$. We relate development of these structures to the gradual formation of the Kolmogorov energy spectrum $E_{k}\propto\, k^{-5/3}$, which we observe in a fully inviscid system. With the spectral analysis we demonstrate that the energy transfer to small scales is performed through the pancake structures, which accumulate in the Kolmogorov interval of scales and evolve according to the scaling law $\omega_{\max}
 \propto \ell^{-2/3}$ for the local vorticity maximums $\omega_{\max}$ and the transverse pancake scales $\ell$.
\end{abstract}

\maketitle

%-------------------------------------------------------------
%-------------------------------------------------------------

\section{Introduction}

The problem of whether incompressible 3D Euler equations develop a singularity in 
finite time (usually termed as blowup or collapse) from smooth initial data of finite energy is 
one of the most long-standing open 
questions in fluid dynamics and applied 
mathematics~\cite{leray1934mouvement,onsager1949statistical,saffman1981dynamics,constantin2007euler,gibbon2008three2}. In physics, formation of 
a collapse is considered as the most effective mechanism enhancing the energy dissipation of regular motion.  
For example, dissipation of 
oceanic surface waves takes significantly less time than its estimate due 
to viscosity~\cite{phillips1959scattering,hasselmann1968weak,kuznetsov1996wave}. The difference is attributed to formation of 
white caps on wave crests and subsequent breaking of waves, which is the manifestation of collapse for surface waves. 
For developed hydrodynamic turbulence, this process is connected with the energy transfer from large (forced) 
scales to small (viscous) scales, where the energy eventually dissipates. 
In the Kolmogorov-Obukhov theory~\cite{kolmogorov1941local,obukhov1941spectral}, 
the velocity fluctuations $\delta v$ at intermediate spatial 
scales $\ell$ obey the power-law  
$\langle|\delta v|\rangle\propto \varepsilon^{1/3} \ell^{1/3}$, 
where $\varepsilon$ is the mean energy flux from large to small scales. This formula can be 
easily obtained from the dimensional analysis, see e.g.~\cite{landau2013fluid,frisch1999turbulence}. 
According to the same dimensional arguments, the fluctuations of vorticity field $\omegabold = \mathrm{rot}\,\mathbf{v}$ diverge at small scales $\ell \to 0$ as $\langle|\delta\omega|\rangle\propto \varepsilon^{1/3} \ell^{-2/3}$, while the time of the 
energy transfer from the energy-contained scale $L$ to the viscous ones is finite and estimated as  
$T\sim  {L^{2/3}}{\varepsilon^{-1/3}}$.
Thus, formation of the Kolmogorov spectrum may be regarded to collapses of vorticity field; such mechanism can be observed in simplified (shell) 
models of turbulence~\cite{mailybaev2013blowup,mailybaev2012computation}. However, finite-time singularities 
are not necessary for the Kolmogorov turbulent spectra, since in case of the finite Reynolds numbers the exponential growth of vorticity is sufficient for energy to reach viscous scales in finite time~\cite{orlandipirozzoli2010}.
See also~\cite{holm2002transient,holm2007} for numerical studies on early stages of turbulent spectra formation. 

There are several blowup and no-blowup criteria for the inviscid flows, which are useful both in analytical studies and numerical simulations. The widely used criterion is due to the Beale--Kato--Majda theorem~\cite{beale1984remarks}, which states 
that the time integral of maximum vorticity must explode at a singular point. 
Several criteria, which  also use the direction of vorticity, are developed by 
Constantin \textit{et al.} \cite{constantin1996geometric}, 
Deng  \textit{et al.} \cite{deng2006improved,deng2005geometric} 
and Chae \cite{chae2007finite}. See also \cite{chae2008incompressible,hou2009blow,gibbon2013dynamics} for other regularity criteria. The blowup scenario based on the vortex lines breaking (or overturning) was analyzed in~\cite{kuznetsov2000collapse} in the framework of the integrable incompressible hydrodynamic model with the Hamiltonian $\int |{\bf \omega}|d{\bf r}$ and the same symplectic operator \cite{kuznetsov1980topological} as for the 3D Euler equations (such unusual Hamiltonian can be obtained from the 3D Euler equations in the so-called local induction approximation). Formation of a singularity in the framework of renormalization group formalism was discussed in~\cite{greene1997evidence,greene2000stability,mailybaev2012renormalization,mailybaev2012renormalizationB}. 

Possible formation of a singularity in incompressible 3D Euler equations has been extensively studied in the 
past decades with direct numerical simulations, which mainly refer to the flow in a box with periodic boundary 
conditions. Below we provide a short review of these numerical results; see also a brief but extensive account in~\cite{gibbon2008three}, as well as \cite{luo2013potentially,larios2014global} for the studies of blowup triggered by the boundary.  

In 1992, Brachet \textit{et al.} \cite{brachet1992numerical} using spectral methods studied 
periodic flows with random initial conditions on $256^3$ grids and also the Taylor--Green vortex on $864^3$ grids, 
and found the energy spectrum well-approximated as 
$E_{k}(t)=ck^{-n(t)}e^{-2\delta(t)k}$.
The exponent $\delta(t)$ clearly demonstrated the exponential decay with time, $\delta(t) \propto e^{-t/T}$. Maximum of vorticity was growing almost exponentially in time with the total increase by the factor of 6 for both initial conditions. The regions of high vorticity represented pancake-like structures in physical space, which were compressed in one direction while their sizes in other two directions did not change considerably. 
Such tendency towards the vortex sheets leads to depletion of nonlinearity and prevents the formation of a finite-time singularity, see the related discussion in~\cite{pumir1990collapsing,ohkitani2008geometrical,kerr2005collapse}.
Thus, further numerical studies were mainly focused on specific initial conditions providing enhanced vorticity growth~\cite{kerr2005collapse}, e.g., antiparallel or orthogonal vortices. Initial conditions were usually chosen to be symmetric, as this required less computational resources. 

In 1993, Kerr \cite{kerr1993evidence} analyzed the interaction of two perturbed antiparallel vortex tubes 
on grids of up to $512\times256\times128$ using the Chebyshev method. The results were interpreted in favor of 
the blowup, 
$\max|\omegabold|\sim (t_0-t)^{-1}$, with an increase of the vorticity maximum by the factor of 24. 
In the subsequent publication~\cite{kerr2005velocity}, the two characteristic length scales of the singularity 
were identified as $\rho \sim (t_0-t)$ and $R \sim (t_0-t)^{1/2}$. However, recent numerical simulations performed 
by Hou and Li~\cite{hou2007computing,hou2009blow} with the pseudo-spectral method on the $1536\times 1024\times 3072$ 
grid questioned the blowup behavior for these initial conditions. The solution was 
extended beyond the earlier estimated blowup time, showing that  
the maximum vorticity evolution is slower than doubly exponential. 
Later the analysis was reconsidered by Bustamante and Kerr~\cite{bustamante20083d}, 
suggesting the hypothesis of 
the vorticity growth as $\max|\omegabold|\sim (t_0-t)^{-\gamma}$ for $\gamma > 1$, 
and then by Kerr~\cite{kerr2013bounds} concluding with the double-exponential growth.

In 1998, Grauer \textit{et al.}~\cite{grauer1998adaptive} used the adaptive mesh refinement technique 
claimed to be equivalent to non-adaptive $2048^3$ grids. The authors achieved maximum vorticity increase of 
about 10 times and interpreted their results in favor of the blowup hypothesis, $\max|\omegabold|\sim (t_0-t)^{-1}$. Similar conclusions were drawn in 2012 by Orlandi~\textit{et al.}~\cite{orlandi2012vortex} for two colliding Lamb dipoles on grids up to $1536^3$. However, in both cases \cite{grauer1998adaptive} and \cite{orlandi2012vortex} the initial conditions had a singularity of vorticity derivatives.

Several studies considered the Pelz--Kida initial flow of high symmetry with the indication of blowup behavior reported in earlier works~\cite{Kida1985,BoratavPelz1994,pelz2001symmetry}. In 2008, Grafke \textit{et al.} \cite{grafke2008numerical} performed comparison of different spectral and real-space numerical methods with the Pelz--Kida like initial flow. The authors achieved the increase of maximum vorticity by about 2.5 times for $512^3$ grids and about 3.5 times for $1024^3$ grids before the methods started to diverge noticeably. Then the adaptive mesh refinement simulations were carried out with the effective resolution of $4096^3$ and vorticity maximum increase of about 6 times. These results, also confirmed by Hou and Li~\cite{hou2008blowup}, demonstrated no tendency toward blowup at the times predicted earlier. It was also noted that the vorticity increased exponentially with time on the Lagrangian trajectory, which ended at the maximum vorticity point at the end of the simulations. 

In the present paper, we analyze 
numerically the 3D Euler equations for rather generic large 
scale initial conditions in a periodic box, i.e., focusing on typical development of vorticity 
in inviscid incompressible flows. 
For numerical simulations, we use the pseudo-spectral method on adaptive rectangular grid. 
Our goal is the systematic study of high vorticity structures, including the maximum vorticity region 
and other local phenomena. 
We identify these structures by looking at local maximums of the vorticity modulus.
We show that a multitude of high vorticity structures appear and have the form of pancakes, i.e., 
thin vorticity sheets compressing exponentially in time in 
a self-similar way for the pancake transverse direction. 
Unlike the pancake model proposed in \cite{brachet1992numerical}, this self-similar dependence is governed by two different exponents  for the pancake compression and the vorticity growth; we suggest how the pancake model \cite{brachet1992numerical} can be modified to capture the observed behavior. Due to the exponential vorticity growth, no tendency toward the blowup is observed.

Our main result is the demonstration of close relation between the collective evolution of pancake 
vorticity structures and the formation of Kolmogorov turbulent spectra. 
Our simulations show that the Kolmogorov energy spectrum starts to form in the fully inviscid system, 
where numerical error is kept very small 
(the flow is not affected by small-scale finite grid effects as opposed 
to \cite{cichowlas2005effective}). With the analysis of local maximums 
we argue that the Kolmogorov spectrum is attributed to the pancake structures, 
which accumulate in the same interval of scales and
evolve according to the scaling law for the local vorticity maximums as
$\omega_{\max} \propto \ell^{-2/3}$, where $\ell$ is the pancake thickness.  
Though our conclusions are limited by the numerical resources that allow the Kolmogorov interval of about one decade, our results provide a new insight on the development of high vorticity structures in relation with the turbulent spectra, 
and indicate the importance of further study in this direction.  
In addition to these results, in our next paper we will present numerical simulations performed in the vortex lines representation variables. The vortex lines representation developed by Kuznetsov and Ruban \cite{kuznetsov1998hamiltonian, kuznetsov2000collapse, kuznetsov2000hamiltonian} describes the flow from the point of view of the moving vortex lines, which are compressible. This representation helps in understanding some of the numerical observations presented in the current paper, including the scaling law $\omega_{\max} \propto \ell^{-2/3}$ between the local vorticity maximums $\omega_{\max}$ and the transverse pancake scales $\ell$.

The paper is organized as follows. Section~\ref{NumMet} describes the numerical method. In Section~\ref{SecGlob} we study the pancake structure near the global vorticity maximum, while in Section~\ref{SecLoc} we focus on the statistics of local maximums. Section~\ref{SecKolm} discusses the numerical observation of the Kolmogorov spectrum and its relation to extreme vorticity structures. The final Section~\ref{SecConcl} contains conclusions. The Appendices~\ref{App:A} and \ref{App:B} contain the initial conditions and 
the results of the second simulation.

%-------------------------------------------------------------
%-------------------------------------------------------------

\section{Numerical method} 
\label{NumMet}

The Euler equations describing dynamics of ideal incompressible fluid of unit density in three-dimensional space are 
\begin{equation}\label{Euler1}
\frac{\partial\mathbf{v}}{\partial t}+(\mathbf{v}\cdot \nabla)\mathbf{v} 
= -\nabla p, \quad \mathrm{div}\, \mathbf{v}=0,
\end{equation}
where the velocity field $\mathbf{v}=(v_{x},v_{y},v_{z})$ and pressure $p$ are smooth functions of space coordinates $\mathbf{r}=(x,y,z)$ and time $t$. We consider solutions of Eq. (\ref{Euler1}) in the box $\mathbf{r} \in [-\pi, \pi]^3$ 
with periodic boundary conditions. For numerical simulations, we use the formulation of Euler equations in terms of 
vorticity (also known as Helmholtz's vorticity equations),
\begin{equation}\label{Euler2}
\frac{\partial\omegabold}{\partial t}=\mathrm{rot}\,(\mathbf{v}\times \omegabold),\quad
\mathbf{v} = \mathrm{rot}^{-1}\omegabold.
\end{equation}
Assuming the vanishing average velocity $\int \mathbf{v} d^3\mathbf{r} = 0$, the inverse of the rotor operator in Eq.~(\ref{Euler2}) is uniquely defined and has the form 
\begin{equation}\label{vorticity_inverse}
\mathbf{v}(\mathbf{k})=\frac{i\mathbf{k}\times\omegabold(\mathbf{k})}{k^{2}}
\quad \textrm{for} \quad k = \|\mathbf{k}\| \ne 0; \quad \mathbf{v}(\mathbf{0}) = \mathbf{0},
\end{equation}
for the Fourier transformed velocity $\mathbf{v}(\mathbf{k})$ and vorticity $\omegabold(\mathbf{k})$. The 
wavevector $\mathbf{k} = (k_{x},k_{y},k_{z})$ has integer components for the periodic box, $\mathbf{r} \in [-\pi, \pi]^3$. Note that the transformation (\ref{vorticity_inverse}) automatically satisfies the incompressibility condition, $\mathbf{k}\cdot\mathbf{v}(\mathbf{k}) = 0$.

\subsection{Adaptive scheme and initial conditions}

We solve the system (\ref{Euler2}) numerically using the pseudo-spectral method
in the adaptive rectangular grid, which is uniform along each coordinate. 
The number of nodes $N_{x}$, $N_{y}$ and $N_{z}$ in each direction is adapted independently, as explained below. To avoid the so-called bottle-neck instability we perform the filtering in Fourier space with the cut-off function~\cite{hou2007computing}
\begin{eqnarray}\label{23rule}
\rho(\mathbf{k}) = \exp\bigg(-36\bigg[\bigg(\frac{k_{x}}{K_{x}}\bigg)^{36} + \bigg(\frac{k_{y}}{K_{y}}\bigg)^{36} + \bigg(\frac{k_{z}}{K_{z}}\bigg)^{36}\bigg]\bigg).
\end{eqnarray}
Here $K_{j}=N_{j}/2$ are the maximum wavenumbers  along directions $j=x,y,z$, i.e., $k_{j} \in [-K_{j}, K_{j}]$. 
Function (\ref{23rule}) cuts off 
approximately 20\% of the spectrum at the edges of the spectral band in 
each direction. We use the fourth-order Runge--Kutta method with the adaptive time stepping, which is implemented according to the CFL stability criterion with the Courant number $0.5$.

%----------------------------------------

\begin{figure}[t]
\centering
\includegraphics[width=8cm]{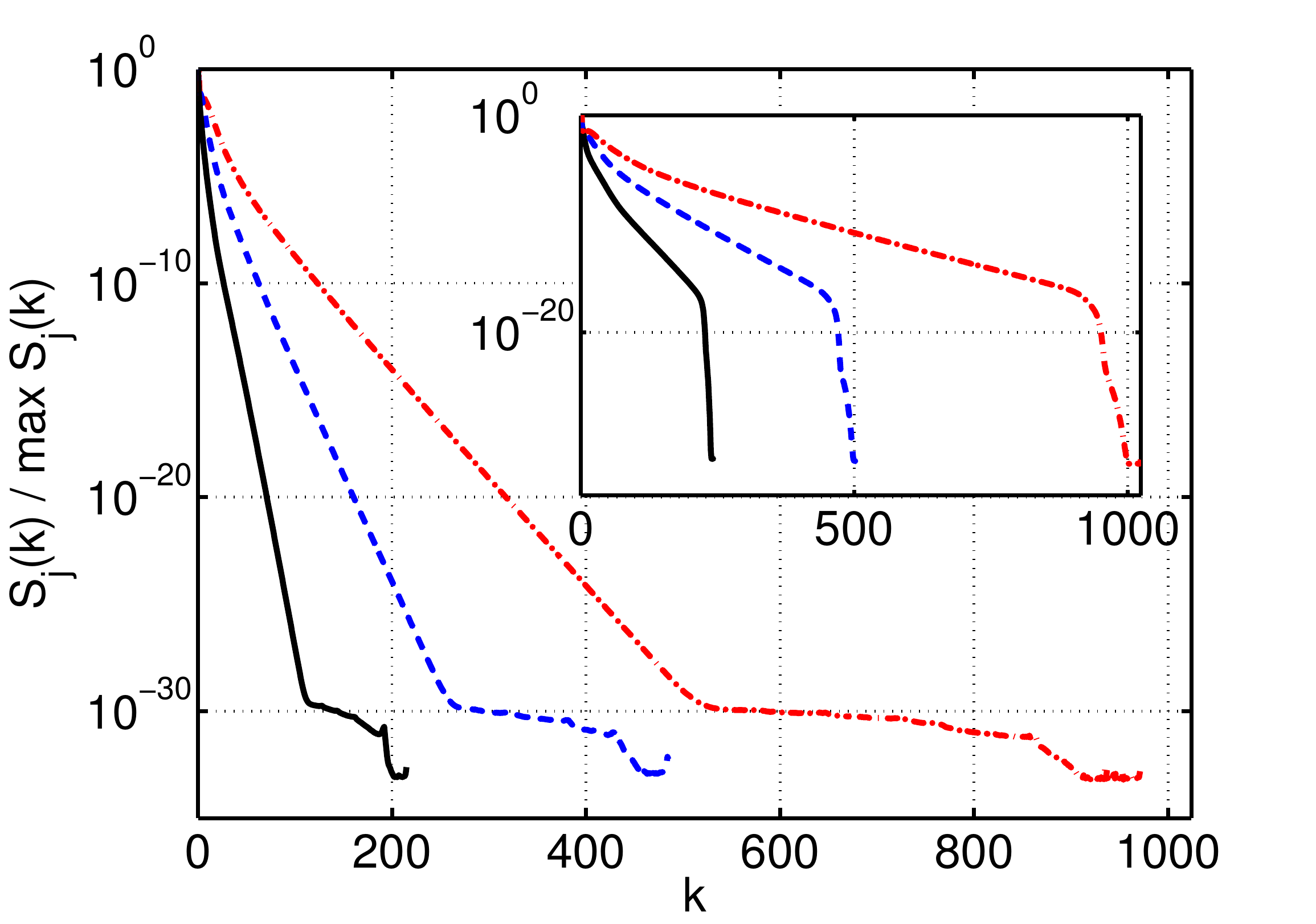}
\caption{{\it (Color on-line)} Normalized spectra of the vorticity field $S_{j}(k)/\max_k S_{j}(k)$ along the $j$-axis ($x$ solid black, $y$ dashed blue, $z$ dash-dot red) for the initial data $I_1$ (see Eq. (\ref{IC1}) 
and Tab.~\ref{tab:M1} in the Appendix~\ref{App:A}) at $t=5$ and at the final time $t=6.89$ (inset).}
\label{fig:Sxyz}
\end{figure}

%----------------------------------------

For optimal performance, we start simulations with the $128^{3}$ grid, which is appropriate for large-scale initial conditions. 
The adaptive scheme is controlled by the functions 
\begin{equation}
S_{j}(k) = \int |\omegabold(\mathbf{p})|^{2}\delta(|p_{j}|-k)\,d^{3}\mathbf{p}, \quad j=x,y,z,
\label{eq_S}
\end{equation}
which describe the enstrophy spectrum along each axis.
As shown in Fig.~\ref{fig:Sxyz}, these functions 
have breakpoints approximately at the level of $\sim\!10^{-30}$, which corresponds to numerical noise. 
During the simulation, these breakpoints move to larger $k$ reflecting the excitation of higher harmonics. 
We stop the simulation every time when one of the breakpoints reaches the value of $2K_{j}/3$ 
for the corresponding axis, and then continue with the refined grid which has increased number of points $N_{j}$ 
along the axis $j$. The vorticity field is interpolated to the new grid using the Fourier interpolation procedure, which has an error comparable to round-off. The simulations carried out in this way are not affected by the aliasing errors. After the total 
number of nodes $N_{x}N_{y}N_{z}$ reaches $1024^{3}$, we continue the simulation 
with the fixed grid and stop when any of the values $S_{j}(2K_{j}/3)$ reaches $10^{-13}\max_k S_{j}(k)$, 
see Fig.~\ref{fig:Sxyz}(inset). 

We tested our simulations with smaller time steps, different harmonics filtering, including the standard 2/3 cut-off 
rule \cite{hou2007computing}, and found negligible difference in the results. 
The simulations conserve the total energy $E = \frac{1}{2}\int |\mathbf{v}|^{2}\,d\mathbf{r}$ 
and helicity $\Omega = \int (\mathbf{v}\cdot \omegabold)\,d\mathbf{r}$
with a relative error smaller than $10^{-11}$. 
We also compared our simulations performed with different limitations for the maximal total number of nodes, and found that the results perfectly converge. For instance, in the time interval $4.13 \le t \le 5.83$, where the simulations with $N_{x}N_{y}N_{z}\le 512^{3}$ and $N_{x}N_{y}N_{z}\le 1024^{3}$ run on different grids, the relative difference for the maximum vorticity is less than $10^{-3}$. The main source of this difference is attributed to the fact that different nodes of the grids are used for approximating the maximum. Note that the high accuracy used in our simulations is crucial for the detailed analysis of extreme structures in the flow, 
which requires very accurate computation of higher-order derivatives near local maximums of the vorticity. For less accurate simulations, numerical errors in these regions increase and the simulation results become unsatisfactory for our purposes. 

We consider initial conditions represented  
in the form of Fourier series 
\begin{equation}
t = 0:\quad \omegabold(\mathbf{r}) =  \sum_{\mathbf{h}} 
\left[\mathbf{A}_\mathbf{h}\cos(\mathbf{h}\cdot\mathbf{r})
+\mathbf{B}_\mathbf{h}\sin(\mathbf{h}\cdot\mathbf{r})\right],
\label{IC1}
\end{equation}
where $\mathbf{h} = (h_x,h_y,h_z)$ is a vector with integer components. Since $\mathrm{div}\,\omegabold = 0$, the real vectors $\mathbf{A}_\mathbf{h}$ and $\mathbf{B}_\mathbf{h}$ must satisfy the orthogonality conditions, $\mathbf{h}\cdot\mathbf{A}_\mathbf{h} = \mathbf{h}\cdot\mathbf{B}_\mathbf{h} = 0$. We fix the vectors $\mathbf{A}_{(0,0,1)}=(0,1,0)$ and $\mathbf{B}_{(0,0,1)}=(1,0,0)$, and choose the other coefficients as random numbers with zero mean and variance $\sigma_{\mathbf{h}}^2 \sim \exp(-|\mathbf{h}|^2)$. Thus, the initial conditions represent the large-scale vorticity field given by the shear flow  (degenerate ABC flow)
\begin{equation}
\omega_{x}=\sin(z),\quad \omega_{y}=\cos(z),\quad \omega_{z}=0,
\label{eq_ABC}
\end{equation}
which is the exact stationary solution of the 3D Euler equations, with a random perturbation. 

A number of initial conditions were tested with the purpose of choosing good candidates for the final high precision simulations. We selected the two initial conditions with better performance for the global vorticity maximum 
and relatively small number of excited harmonics, 
which are denoted as $I_1$ and $I_2$ and summarized in Tabs.~\ref{tab:M1} and \ref{tab:M2} in the Appendix~\ref{App:A} along 
with some simulation information. We focus our analysis on the simulation for the initial condition $I_1$, which 
reaches the time $t=6.89$ on the grid $486\times 1024\times 2048$. The different number of grid points for different directions results from the adaptive scheme, which resolves the anisotropy of vorticity in an optimal way. In the Sections below this simulation will be always assumed unless otherwise stated. The simulation results for the initial condition $I_2$ are summarized in the Appendix~\ref{App:B}. 

\subsection{Analysis of high vorticity structures}
\label{sec:numC}

\begin{figure}[t]
\centering
\includegraphics[width=8cm]{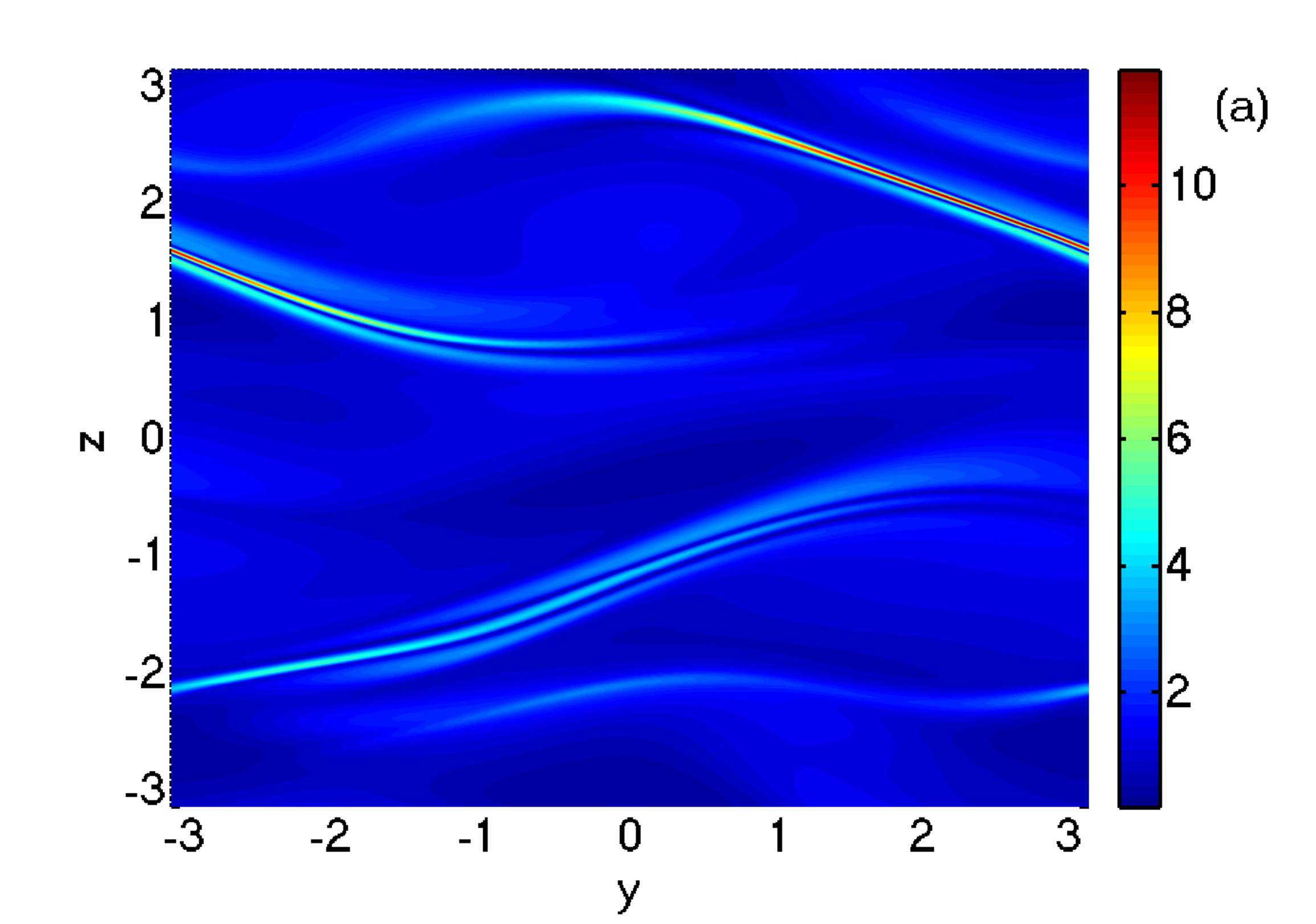}
\includegraphics[width=8cm]{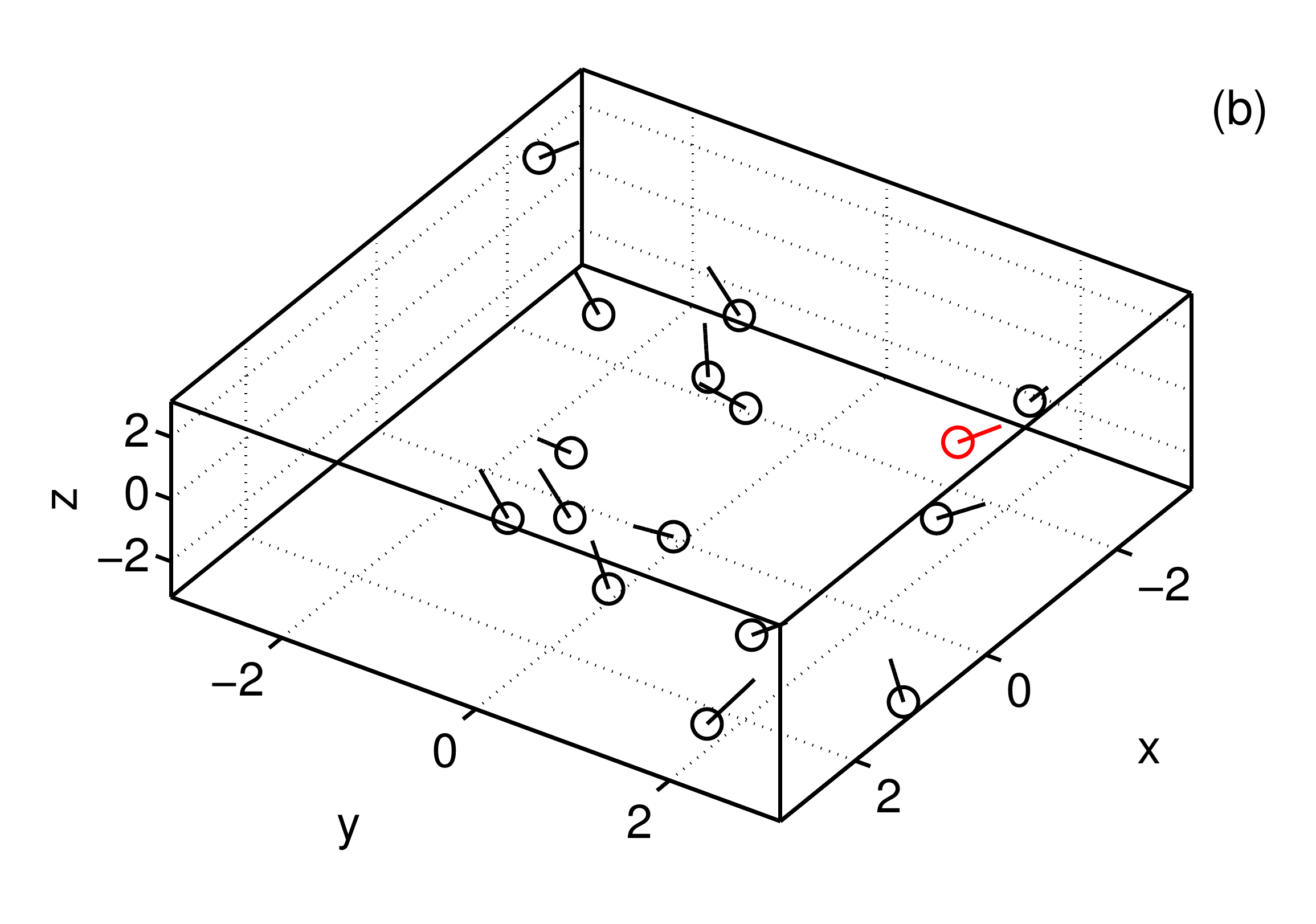}
\caption{{\it (Color on-line)} (a) Vorticity field $|\omegabold|$ at fixed $x=x_{0}$ at the final time $t = 6.89$. Here $x_{0}$ is the $x$-coordinate of the global maximum. (b) Positions (circles) of local maximums of $|\omegabold|$ and directions (lines) of the eigenvectors $\mathbf{w}_1$ normal to the pancake structures, at $t = 6.89$. Red color marks the global maximum.
}
\label{fig5}
\end{figure}

As we demonstrate in the next Sections, the regions of high vorticity are strongly anisotropic and develop in the form 
of pancake-like structures, which are thin in one direction and remain large 
in other two directions, Fig.~\ref{fig5}(a). Conclusions of our paper rely 
significantly on the systematic identification of such structures, which is performed by searching for local maximums 
of the vorticity modulus $|\omegabold(\mathbf{r})|$. The values of vorticity in the pancake are very sensitive to the distance 
from the pancake midplane, while the shifts along this plane have much smaller effects. Thus, 
searching for maximums by a simple comparison of neighboring grid points yields a large number (up to thousands) of ``false'' local maximums. For this reason, we developed 
a three-step numerical procedure for the identification of ``true'' local maximums. 

At each grid point $\mathbf{r}_g$, we compute the gradient vector $\mathbf{g}(\mathbf{r}_{g}) = \nabla|\omegabold|$ and the Hessian matrix $\mathbf{H}(\mathbf{r}_{g}) = [\,\partial^{2}|\omegabold|/\partial x_{i}\partial x_{j}\,]$ using the spectral method. A local maximum is characterized by the vanishing gradient vector and negative definite Hessian matrix, which has three negative eigenvalues $\lambda_1 < \lambda_2 < \lambda_3 < 0$ with the orthonormal eigenvectors $\mathbf{w}_1$, $\mathbf{w}_2$, $\mathbf{w}_3$. As we will demonstrate below for the local maximums of vorticity, $|\lambda_{1}|\gg |\lambda_{2}|\sim|\lambda_{3}|$ and the eigenvector $\mathbf{w}_1$ determines the perpendicular direction to the pancake midplane while the eigenvectors $\mathbf{w}_2$ and $\mathbf{w}_3$ lie in this plane. 
Assuming that the local maximum $\mathbf{r}_m$ lies near $\mathbf{r}_g$, the second-order approximation yields
\begin{equation}\label{Taylor_XYZ_lm_position}
\mathbf{r}_{m} \approx \mathbf{r}_{g} - \mathbf{g}(\mathbf{r}_{g})\mathbf{H}^{-1}(\mathbf{r}_{g}).
\end{equation}
As the first step in our procedure, we check every grid node $\mathbf{r}_g$ to satisfy the following two conditions simultaneously: Eq. (\ref{Taylor_XYZ_lm_position}) yields the point $\mathbf{r}_m$ lying in one of the adjacent grid cells, and the Hessian matrix $\mathbf{H}(\mathbf{r}_{g})$ is negative definite. These conditions determine a cloud of grid points around each local maximum. So, as the second step, we select a single node in each cloud by choosing the node with the smallest distance $|\mathbf{r}_{m}-\mathbf{r}_{g}|$. At larger times some of the pancake structures contain several local maximums of vorticity. Some of these maximums form localized clusters, within which local maximums have close values of vorticity, almost the same eigenvectors $\mathbf{w}_{1}$, and lie approximately in the same plane almost perpendicular to the eigenvectors $\mathbf{w}_{1}$. Thus, such maximums represent close and similar oriented parts of the same pancake. In the third step we determine all such 
clusters of local maximums, and leave only one point with the largest value of vorticity from each cluster.

We expect that the local maximums in the resulting set are associated with different pancake structures or different (distant or differently oriented) parts of the same pancake, Fig.~\ref{fig5}(b). The proposed method allows adequate selection of local maximums of vorticity, producing a little amount of false maximums. We compared its performance on significantly different grids, and found that the resulting sets of local maximums differ by no more than 10-15\%.

%-----------------------------------------------------------------------------------------------------------------------------------

\section{Evolution near the global maximum of vorticity}
\label{SecGlob}

\begin{figure}[t]
\centering
\includegraphics[width=8cm]{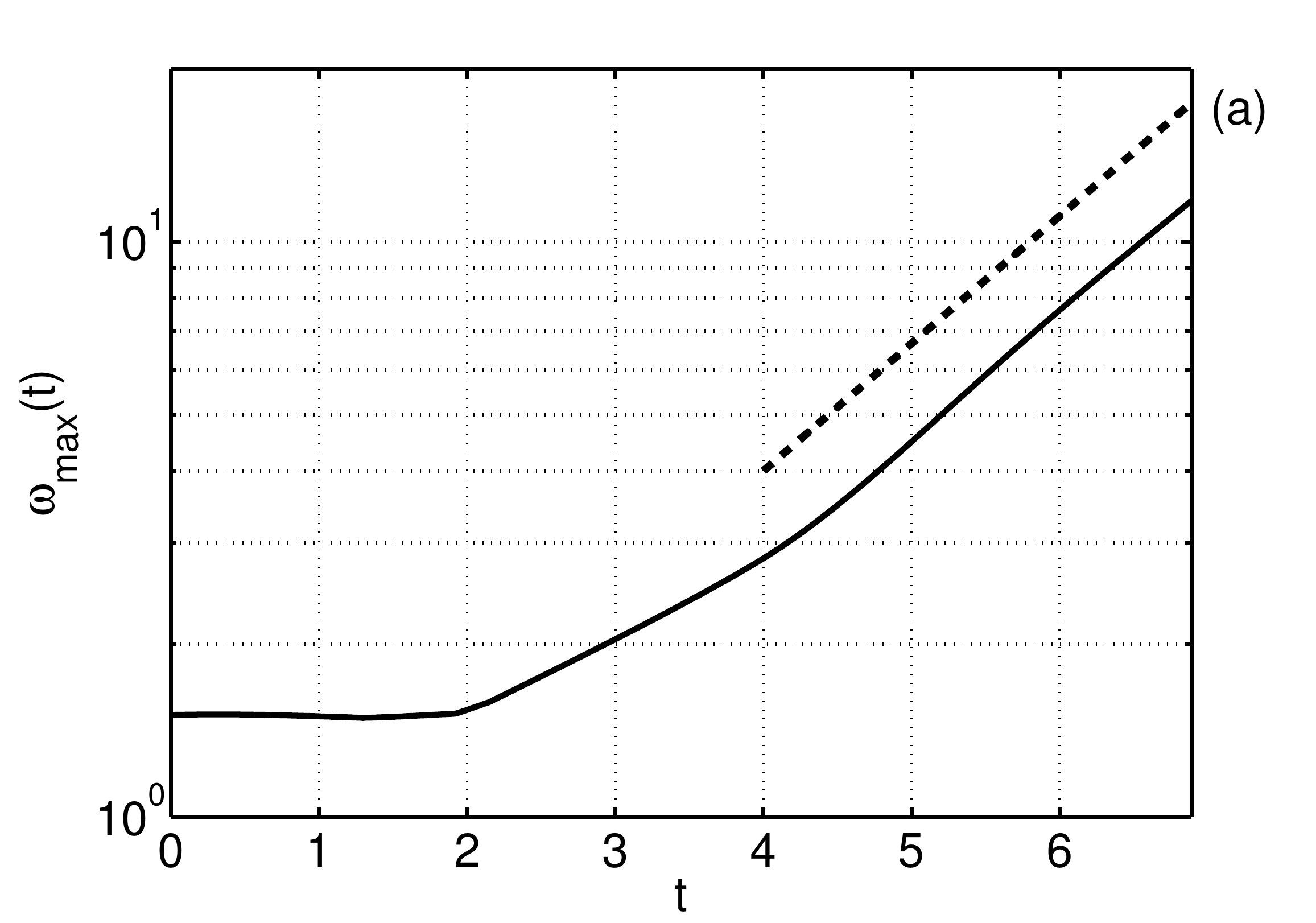}
\includegraphics[width=8cm]{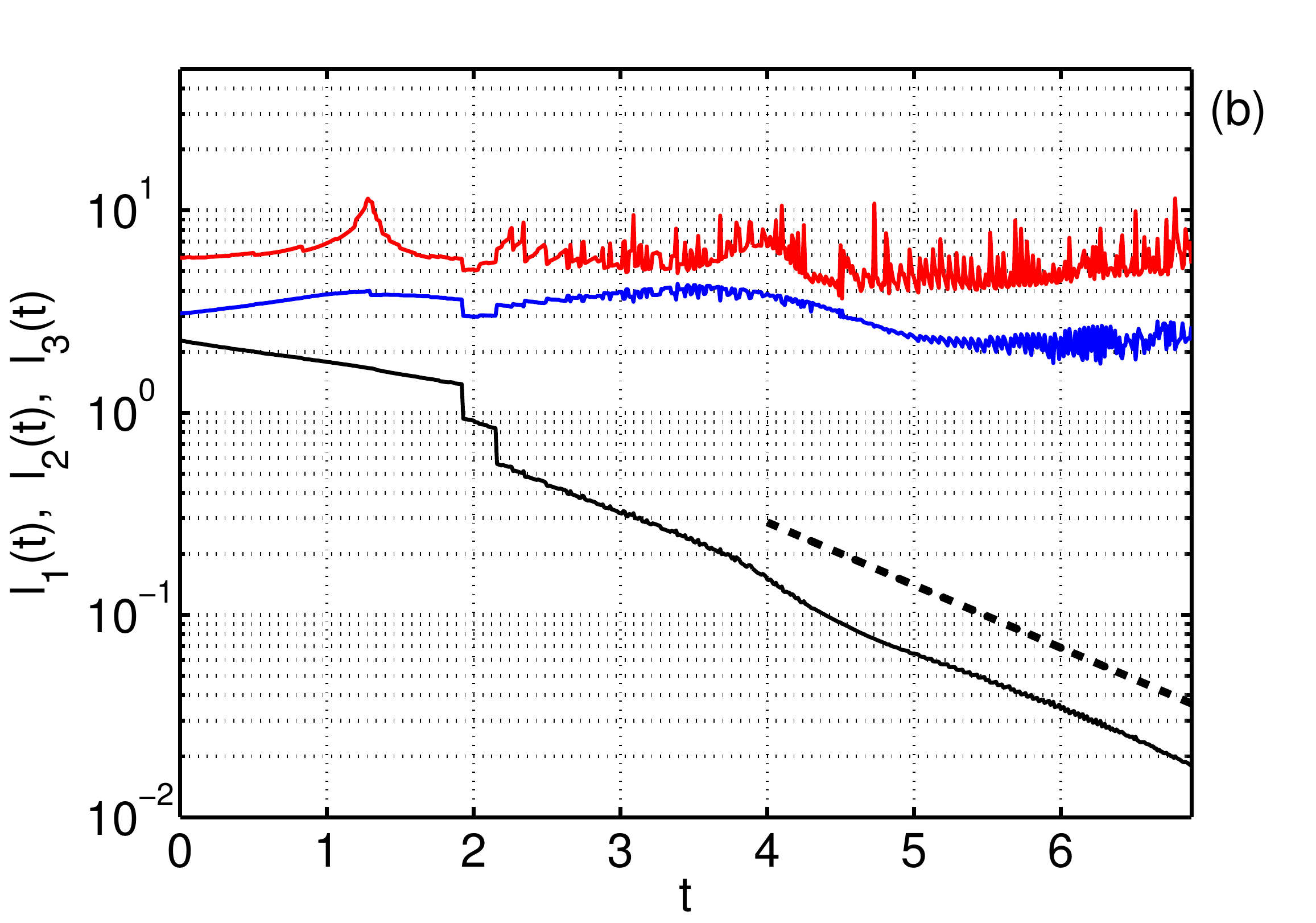}
\caption{{\it (Color on-line)} (a) Evolution of the global vorticity maximum (logarithmic vertical scale). 
The dashed line indicates the exponential slope $\propto\!e^{t/T_{\omega}}$ with characteristic time $T_{\omega} = 2$.
(b) Evolution of characteristic spatial scales for the region of the global vorticity maximum, $\ell_{1}$ (black), $\ell_{2}$ (blue) and $\ell_{3}$ (red). The dashed line indicates the exponential slope $\propto\!e^{-t/T_{\ell}}$ with characteristic time $T_{\ell} = 1.4$. }
\label{fig:evolution_lm_dimensions}
\end{figure}

Numerical studies of singularities in the 3D incompressible Euler equations 
use the evolution of global vorticity maximum $\omega_{\max}(t) = \max|\omegabold(\mathbf{r})|$ as one of the principal tests. 
According to the Beale--Kato--Majda theorem~\cite{beale1984remarks}, unbounded increase of this maximum is necessary 
for the finite-time blowup. In this Section we consider the dynamics of $\omega_{\max}(t)$ and analyze the local 
geometry of the flow. The global vorticity maximum grows with time from the initial value $\omega_{\max}(0)\approx 1.5$ 
to $11.8$ at the finial time of the simulation $t = 6.89$, Fig.~\ref{fig:evolution_lm_dimensions}(a). For $t > 4.5$, 
this growth is well-approximated by the exponential function 
$\omega_{\max}(t)\propto e^{t/T_{\omega}}$ with $T_{\omega}\approx 2$.
Contrary to the vorticity, the velocity maximum $\max|\mathbf{v}(\mathbf{r})|$ does not change 
more than by $10\%$ during the whole simulation. 

\begin{figure}[t]
\centering
\includegraphics[width=8cm]{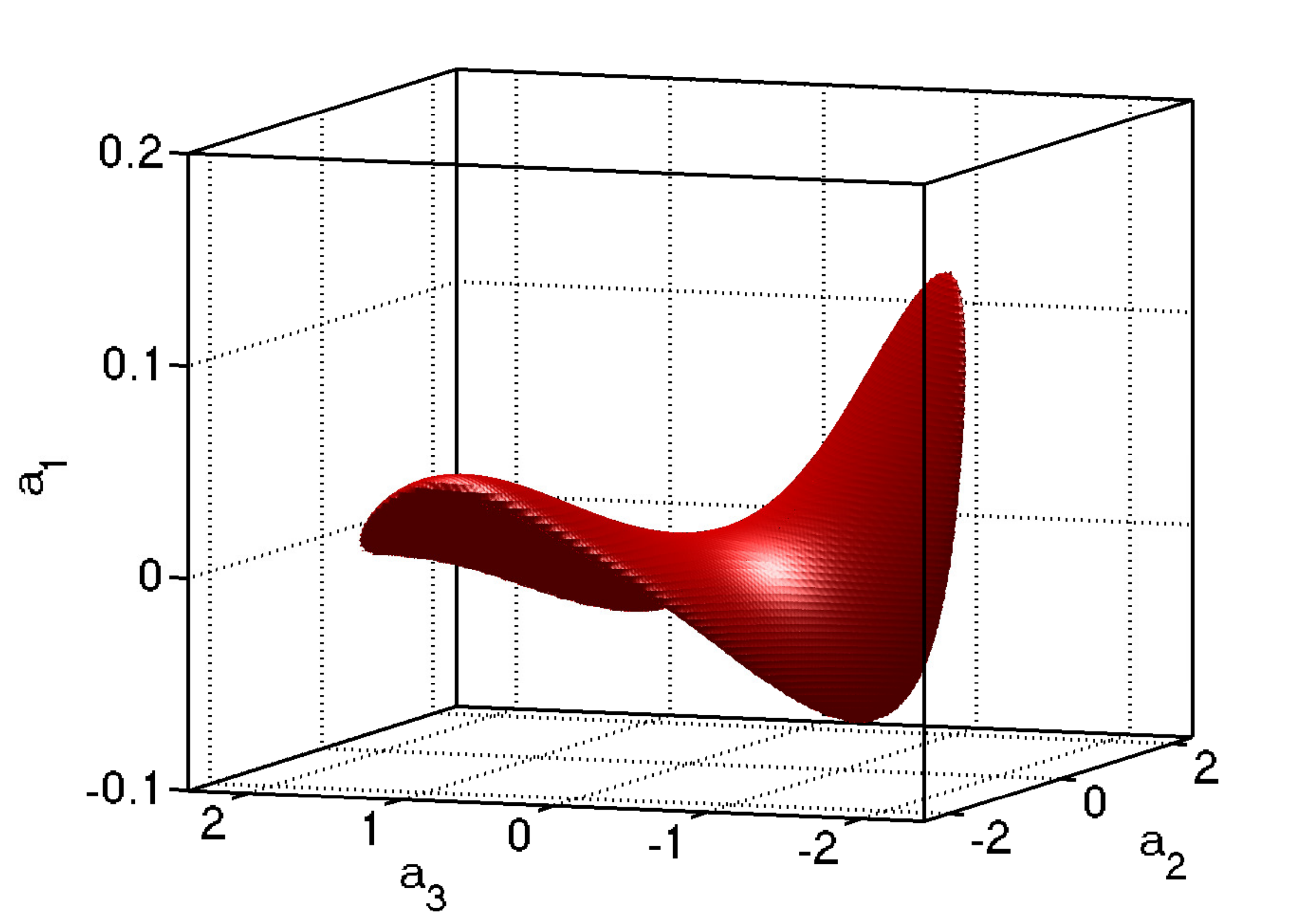}
\caption{{\it (Color on-line)} Isosurface of constant vorticity $|\omegabold|=0.8\,\omega_{\max}$ in the local coordinates $(a_{1},a_{2},a_{3})$ at the final time of the simulation, $t=6.89$. Note much smaller vertical scale.}
\label{fig:evolution_lm_dimensions_iso}
\end{figure}

Local geometry near the vorticity maximum can be studied using the Hessian matrix $\mathbf{H} = [\,\partial^{2}|\omegabold|/\partial x_{i}\partial x_{j}\,]$ computed at the maximum point $\mathbf{r}_m$. This matrix has three negative eigenvalues $\lambda_1 < \lambda_2 < \lambda_3 < 0$ with the orthogonal eigenvectors  $\mathbf{w}_1$, $\mathbf{w}_2$, $\mathbf{w}_3$. Considering the local orthonormal basis as $\mathbf{r}=\mathbf{r}_{m} + a_{1}\mathbf{w}_1 + a_{2}\mathbf{w}_2+ a_{3}\mathbf{w}_3$, the vorticity modulus can be described by the quadratic approximation
\begin{equation}
\frac{|\omegabold(\mathbf{r})|}{\omega_{\max}} =  1
-\left(\frac{a_1}{\ell_1}\right)^2
-\left(\frac{a_2}{\ell_2}\right)^2
-\left(\frac{a_3}{\ell_3}\right)^2
+o(|\mathbf{r}-\mathbf{r}_{m}|^2),\quad
\ell_{j} = \sqrt{\frac{2\omega_{\max}}{-\lambda_{j}}}.
\label{Taylor_A123}
\end{equation}
The quantities $\ell_{1}$, $\ell_{2}$ and $\ell_{3}$ determine the size of high vorticity region 
at fixed time, and their time evolution is shown in Fig.~\ref{fig:evolution_lm_dimensions}(b).
Note that the discontinuities in this figure near $t=2$ correspond to the change of global maximum among different local maximums, 
while the noise in determining $\ell_{2}$ and $\ell_{3}$ for larger times is the result of amplified numerical error 
due to the ill-conditioned Hessian matrix. 
The smallest size decreases exponentially in time, $\ell_1 \propto e^{-t/T_{\ell}}$ with $T_{\ell} \approx 1.4$ for $t > 4.5$, 
while $\ell_2$ and $\ell_3$ remain almost the same. At final time, we have $\ell_1/\ell_{2} \sim \ell_1/\ell_{3} \sim 10^{-2}$, 
which implies that the high-vorticity region represents a very thin pancake structure with the normal vector $\mathbf{w}_{1}$ 
and tangent vectors $\mathbf{w}_{2}$ and $\mathbf{w}_{3}$. This is confirmed in Fig.~\ref{fig:evolution_lm_dimensions_iso} 
(note the much smaller scale of the vertical axis)
showing the numerically computed isosurface $|\omegabold(\mathbf{r})| = 0.8\,\omega_{\max}$ in 
the coordinates $(a_{1},a_{2},a_{3})$. 

Similar pancake structures were observed systematically in~\cite{brachet1992numerical}, where the singularity was analyzed by looking at the analyticity strip of the solution. This method relies on the evolution of the energy spectrum 
\begin{equation}\label{Ek}
E_{k}(t) = \frac{1}{2}\int |\mathbf{v}(\mathbf{k},t)|^{2}\,k^{2}do,
\end{equation}
where $o$ is the spherical angle. Numerically, we find $E_{k}(t)$ as a sum $\frac{1}{2}\sum|\mathbf{v}(\mathbf{k})|^{2}$ over 
all nodes in the spherical shell $k\le |\mathbf{k}|< k+1$. This procedure is simple and yields the result, 
which is very close to 
the direct computation of the integral in Eq.~(\ref{Ek}) with the interpolation of velocity Fourier components on the 
sphere of radius $k$.

\begin{figure}[t]
\centering
\includegraphics[width=8cm]{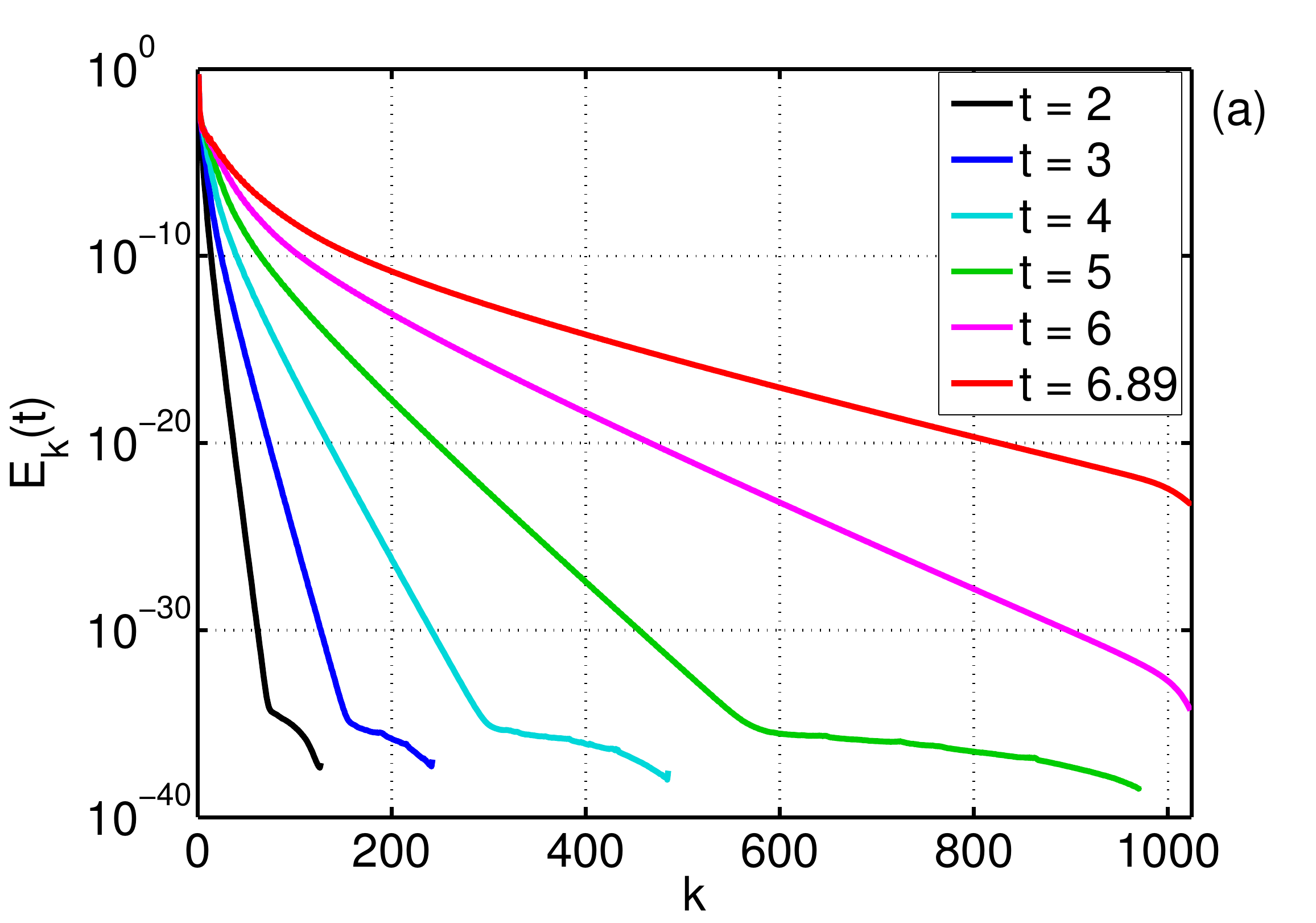}
\includegraphics[width=8cm]{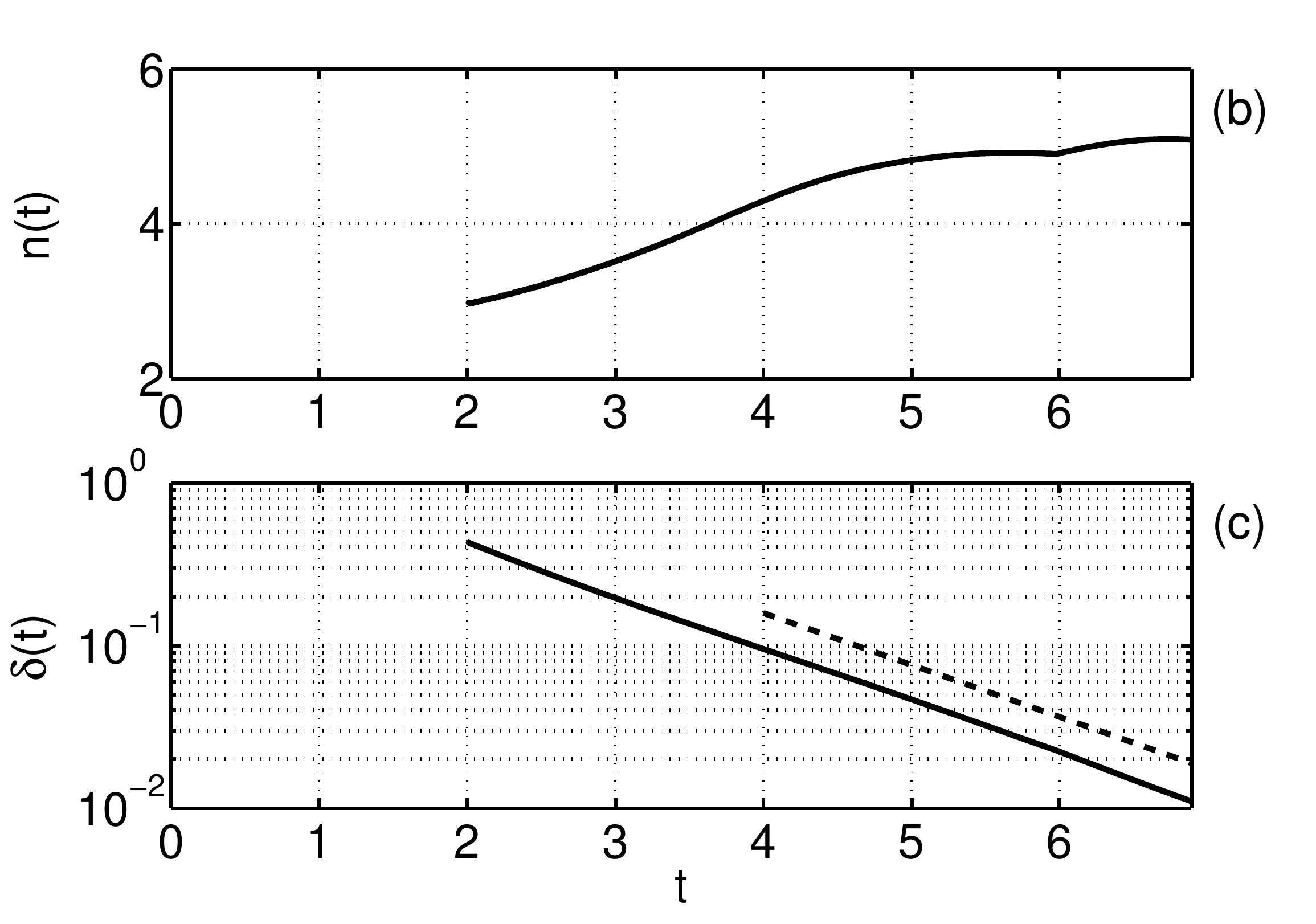}
\caption{{\it (Color on-line)} (a) Energy spectrum $E_{k}(t)$ at different times. Figures (b) and (c) show the exponents $n(t)$ and $\delta(t)$ of the fit $E_{k}(t) \approx c(t)k^{-n(t)}e^{-2\delta(t) k}$. The dashed line in figure (c) indicates the exponential slope $\propto\!e^{-t/T_{\delta}}$ with characteristic time $T_{\delta} = 1.4$.}
\label{fig:Ek}
\end{figure}

For fixed time and sufficiently large wavenumbers, the energy spectrum decays with $k$ exponentially,  
$E_{k}(t) \propto e^{-2\delta(t) k}$, 
possibly with an algebraic prefactor, until the level of $\sim\! 10^{-35}$ corresponding to 
numerical noise, Fig.~\ref{fig:Ek}(a). The exponent $\delta(t)$ can be associated with the width of analyticity strip for the solution extended to complex coordinates $\mathbf{r}$. In \cite{brachet1992numerical}, the spectrum fitting 
$E_{k}(t) \approx c(t)k^{-n(t)}e^{-2\delta(t) k}$ was used and the function $\delta(t)$ showed the exponential decay $\delta(t)\propto e^{-t/T_{\delta}}$ at large times. Our simulations lead to the same conclusions with $T_{\delta}\approx 1.4$, Fig.~\ref{fig:Ek}(c).
Note that $T_{\delta}\approx T_{\ell}$, where $T_{\ell}\approx 1.4$ describes the exponential decay of the pancake width 
(we estimate the absolute numerical accuracy for the time scales $T_{\omega}$, $T_{\ell}$ and $T_{\delta}$ as $\pm 0.1$). 
This relation is natural, since the analyticity strip must be determined by the most extreme event, i.e., by the thinnest part of a pancake at the point of maximum vorticity. 
Note that our results for the algebraic prefactor $n(t)$ differ from \cite{brachet1992numerical}, where $n(t)$ approached $-4$ at late times, Fig.~\ref{fig:Ek}(b).

The following self-similar solution of the Euler equations was suggested in \cite{brachet1992numerical} for the description of the flow structure in a small neighborhood of the pancake,
\begin{equation}
v_1 = -\frac{ a_1}{T},\quad
v_2 = \frac{ a_2}{T},\quad
v_3 = f\left( a_1e^{t/T}\right),\quad
p = -\frac{a_1^2 + a_2^2}{2T^2},
\label{eq1}
\end{equation}
which is written in local coordinates $(a_1, a_2, a_3)$ with the axis $a_1$ perpendicular to the pancake; $v_{j}$ are components of the velocity, $p$ is pressure and $f(a_1)$ is an arbitrary function. The model (\ref{eq1}) leads to the exponential growth $\propto e^{t/T}$ of vorticity $\omegabold = (\omega_{1},\omega_{2},\omega_{3})$ as
\begin{equation}
\omega_{1} = \omega_{3} = 0, \quad \omega_{2} = -e^{t/T}f'(a_{1}e^{t/T})
\label{eq1_1}
\end{equation}
in the pancake, whose thickness decreases exponentially as $\propto e^{-t/T}$ with the same characteristic time $T$. However, our simulations demonstrate that the evolution of the pancake is governed by the two different exponents $e^{t/T_{\omega}}$, $T_{\omega}\approx 2$, and $e^{-t/T_{\ell}}$, $T_{\ell}\approx 1.4$, for the vorticity growth and the pancake compression, respectively. We propose that the flow near the pancake can be approximated as
\begin{equation}
v_1 = -\frac{ a_1}{T_{\ell}},\quad
v_2 = \frac{ a_2}{T_{\ell}},\quad
v_3 = e^{-\alpha t}f\left( a_1e^{t/T_{\ell}}\right),\quad
p = -\frac{a_1^2 + a_2^2}{2T_{\ell}^{2}},
\label{eq2}
\end{equation}
where $\alpha=1/T_\ell-1/T_\omega>0$. The modified model (\ref{eq2}) is not the solution of the Euler equations. However, it satisfies the incompressibility condition $\mathrm{div}\,\mathbf{v}=0$, yields the correct exponents for the maximum vorticity and the pancake thickness
\begin{equation}
\omega_{1} = \omega_{3} = 0, \quad \omega_{2} = -e^{t/T_{\omega}}f'(a_{1}e^{t/T_{\ell}}),
\label{eq2_1}
\end{equation}
and after being substituted into the Euler equations (\ref{Euler1}), ensures the cancellation of the leading-order terms growing exponentially as $\propto e^{t/T_{\omega}}$, while the next-order terms (not canceled) decay as $\propto e^{-\alpha t}$.

\begin{figure}[t]
\centering
\includegraphics[width=8cm]{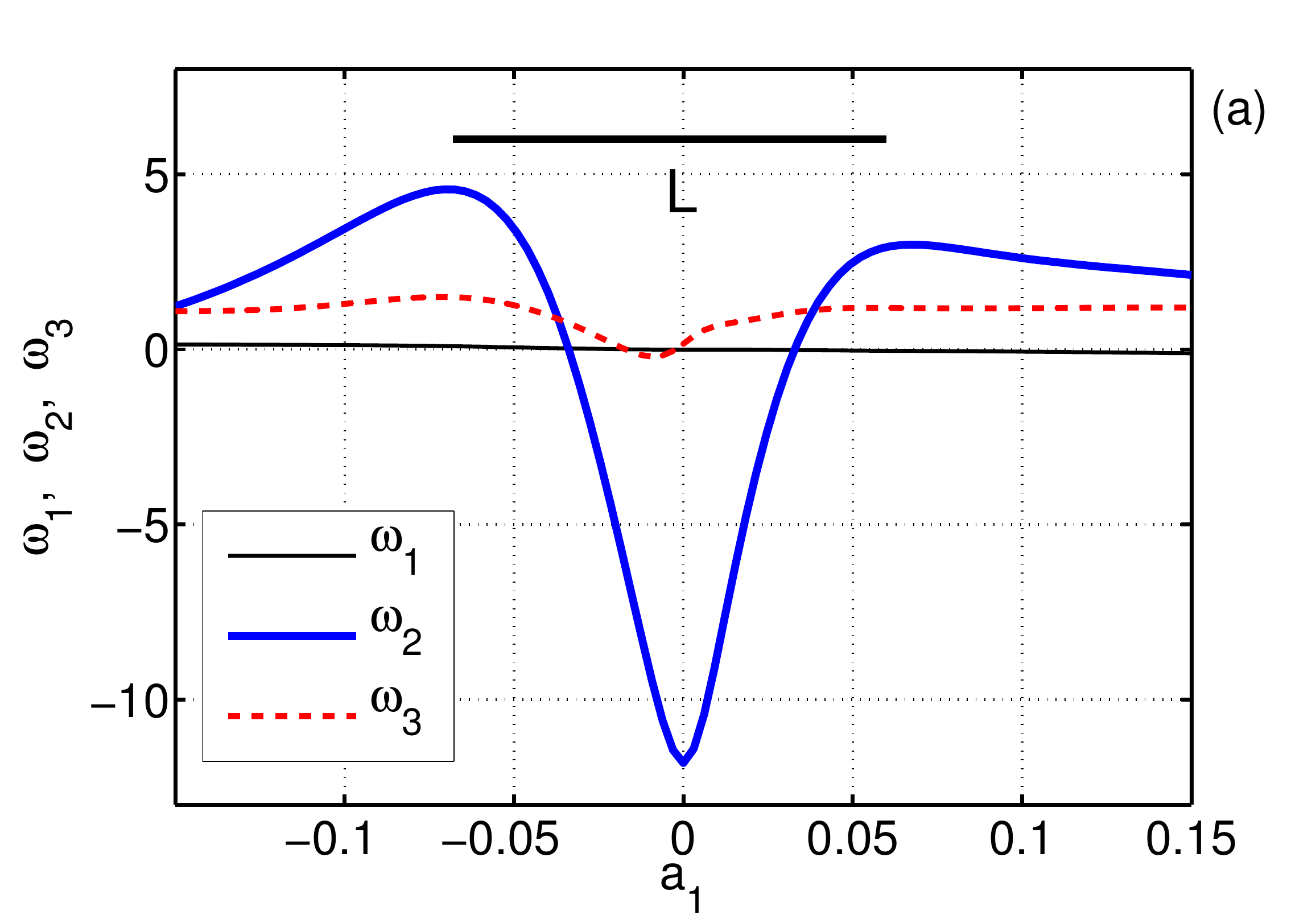}
\includegraphics[width=8cm]{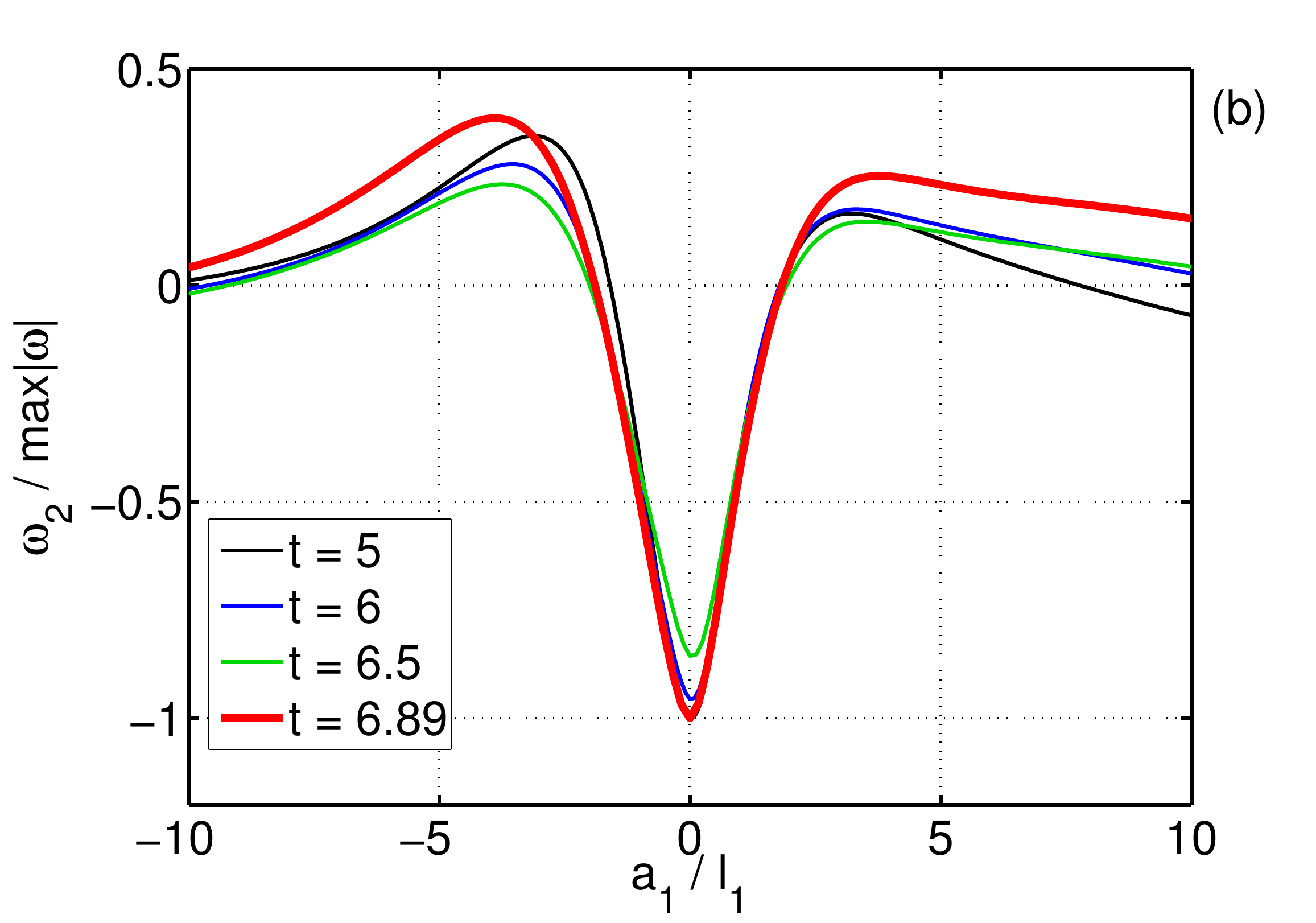}
\caption{{\it (Color on-line)} 
(a) Components of the vorticity vector $\omegabold = \omega_1\mathbf{w}_1+\omega_2\mathbf{w}_2+\omega_3\mathbf{w}_3$ along the axis 
$a_1$ perpendicular to the pancake at $t = 6.89$. (b) Renormalized vorticity component $\omega_2/\omega_{\max}(t)$ vs. 
renormalized coordinate $a_1/\ell_1$ at different times demonstrating self-similarity of the vorticity field.
}
\label{figX}
\end{figure}

Our pancake self-similarity hypothesis agrees well with the simulation results, see Fig.~\ref{figX}. Note that the vorticity vector near the pancake is mostly aligned with the eigenvector $\mathbf{w}_{2}$ (axis $a_2$), in agreement with the model given by Eqs.~(\ref{eq2})-(\ref{eq2_1}). 
One can expect that the pancake (\ref{eq2})-(\ref{eq2_1}) contributes to the energy spectrum in the interval 
of wavenumbers $|k|\lesssim k_{1}$ with 
\begin{equation}
k_{1} = \frac{1}{\ell_1} \sim \frac{2\pi}{L},
\label{eq3}
\end{equation}
where $\ell_1 \propto e^{-t/T_\ell}$ is the characteristic scale 
along the axis $a_1$ in Eq.~(\ref{Taylor_A123}) and 
$L$ estimates the oscillation period of vorticity components in Fig.~\ref{figX}(a); 
this relation can be deduced, e.g., 
by analogy with the Fourier transform of the 
function $\mathrm{sech}(x/\ell_1)$. At final time, the numerical simulation yields $L\approx 0.13$ and $\ell_{1}\approx 0.018$, which is in good agreement with relation (\ref{eq3}).
 
It is worth noting that self-similar model (\ref{eq2})--(\ref{eq2_1}) cannot be interpreted in the context of two-dimensional flow, since vorticity grows exponentially in time, while the vorticity vector is parallel to the pancake plane and is not orthogonal to the velocity. This indicates the importance of third dimension for the pancake dynamics.

The second simulation with the initial condition $I_2$ follows the same scenario 
for the global vorticity maximum and the associated region of high vorticity; see Appendix~\ref{App:B}. 
The corresponding characteristic times 
for the exponential behavior are estimated as $T_{\omega}\approx 2.7$, $T_{\ell}\approx 1.7$ and $T_{\delta}\approx 1.7$. 
Again, the two time scales $T_{\delta}\approx T_{\ell}$ are close, while the scale $T_{\omega}$ is considerably larger. We conclude that the pancake behavior is governed by the two characteristic times controlling 
the exponential growth of vorticity and the exponential decrease of pancake thickness. 

%----------------------------------------

\section{Local maximums of vorticity}
\label{SecLoc}

Figure~\ref{fig5}(a) shows the vorticity distribution for the cross-section passing through the global maximum at final time. 
One can see that the regions of increased vorticity 
have a tendency to form a number of (thin and wide) pancake structures, which is in agreement with earlier simulations 
for generic initial conditions~\cite{brachet1992numerical}. We expect that the structural analysis of these pancakes 
might give insight into the 3D Euler dynamics. However, direct identification of the pancakes is a complicated numerical problem.  
We approach this problem by finding and analyzing local maximums of the vorticity modulus, as described in Section~\ref{sec:numC}.

\begin{figure}[t]
\centering
\includegraphics[width=8cm]{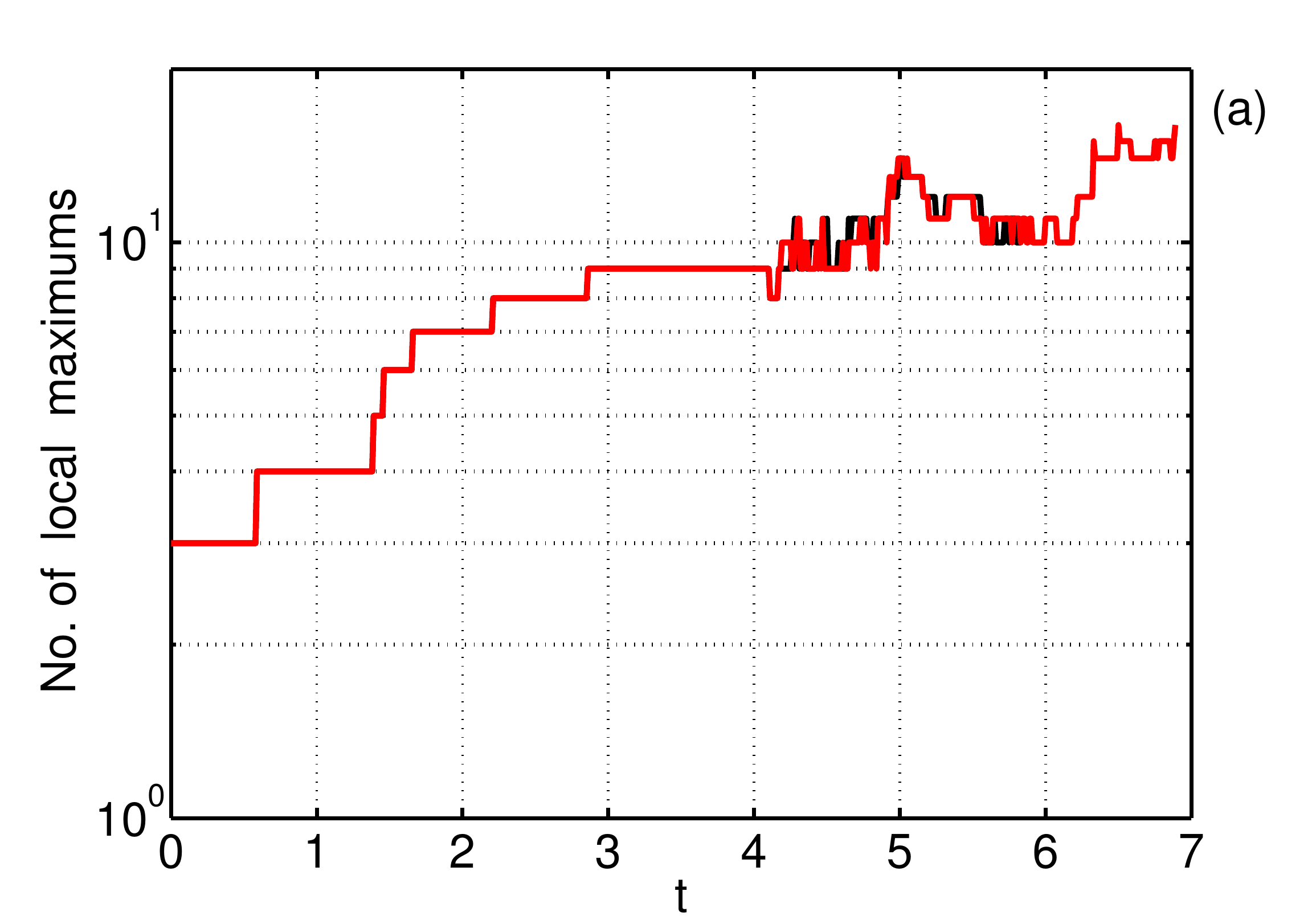}
\includegraphics[width=8cm]{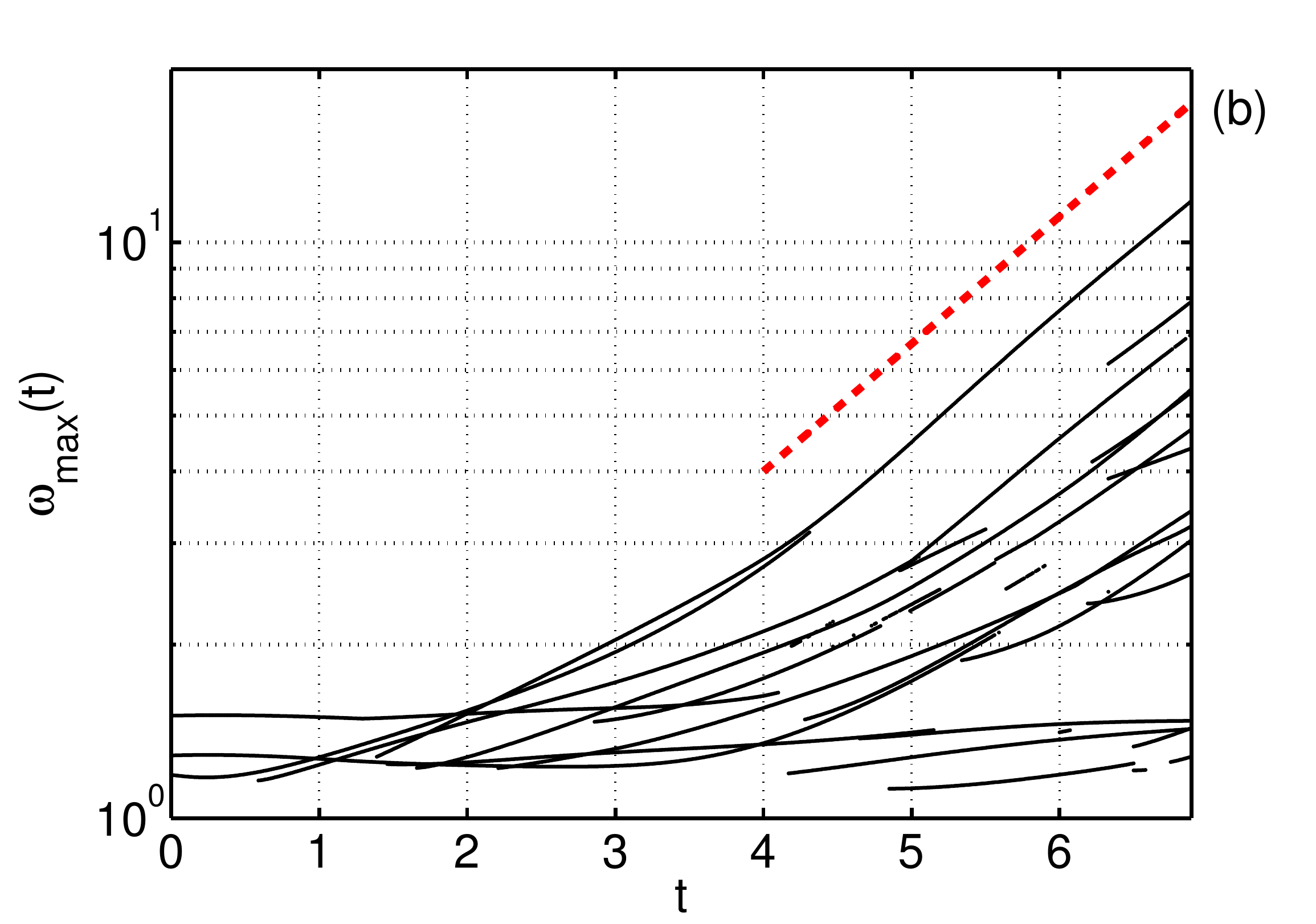}
\caption{{\it (Color on-line)}
(a) Total number of local vorticity maximums versus time obtained for the two simulations corresponding to the 
total number of nodes $N_{x}N_{y}N_{z} \le 512^{3}$ (black, ends at $t=5.83$) and $1024^3$ (red). 
(b) Evolution of local vorticity maximums (logarithmic vertical scale). The dashed red line indicates the exponential slope $\propto e^{t/T_{\omega}}$ with characteristic time $T_{\omega} = 2$.}
\label{fig:W_along_w1}
\end{figure}

Figure~\ref{fig:W_along_w1}(a) shows that the number of local maximums increases with time, from $3$ at $t = 0$ 
to $16$ at the end of the simulation. 
The values of vorticity modulus at maximum points are shown in Fig.~\ref{fig:W_along_w1}(b). The results 
suggest that the vorticity growth tends to be exponential, $\omega_{\max}(t)\propto e^{t/T_{\omega}}$, 
for many of the local maximums with rather close values of the characteristic times $T_{\omega}$. 

\begin{figure}[t]
\centering
\includegraphics[width=8cm]{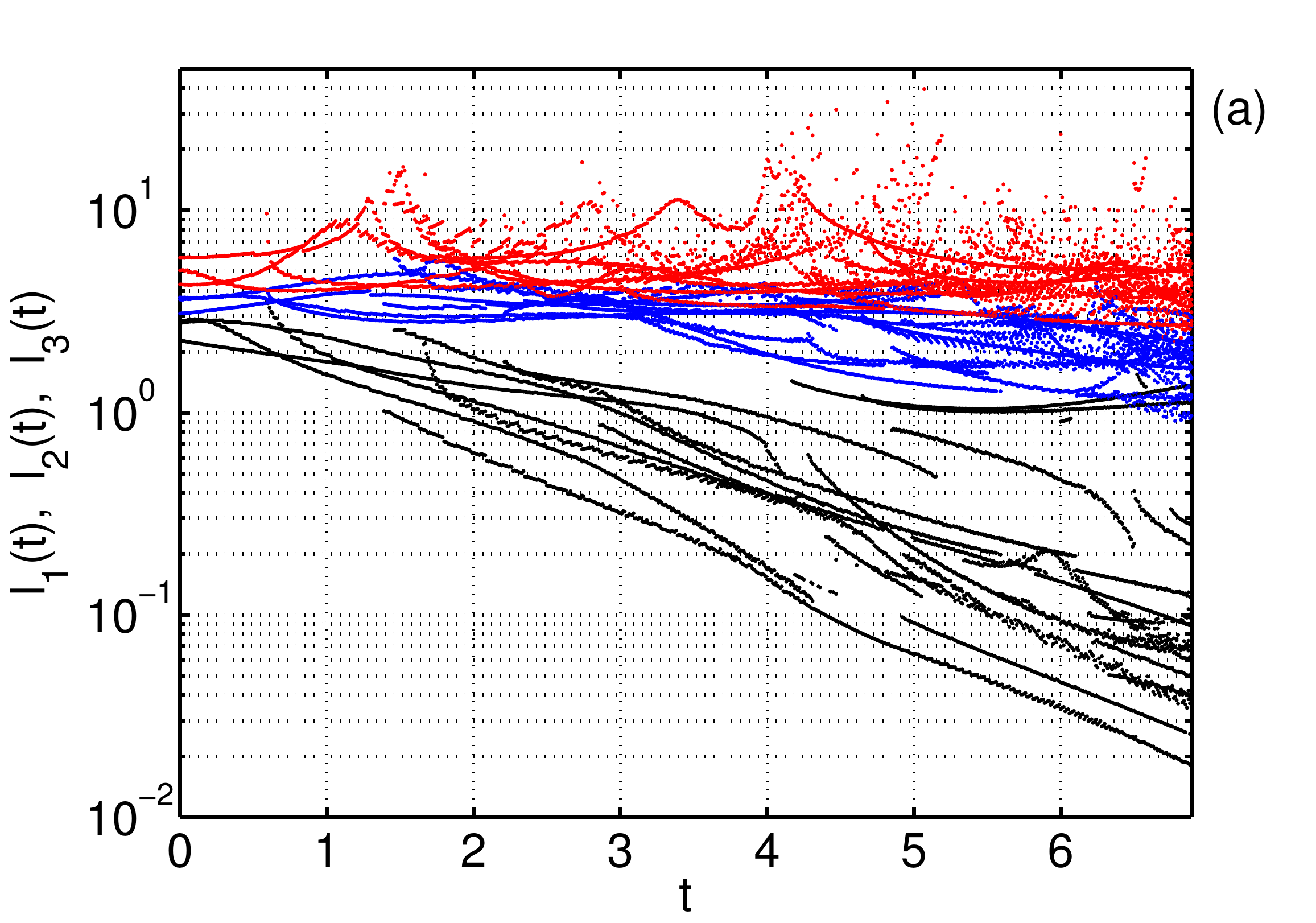}
\includegraphics[width=8cm]{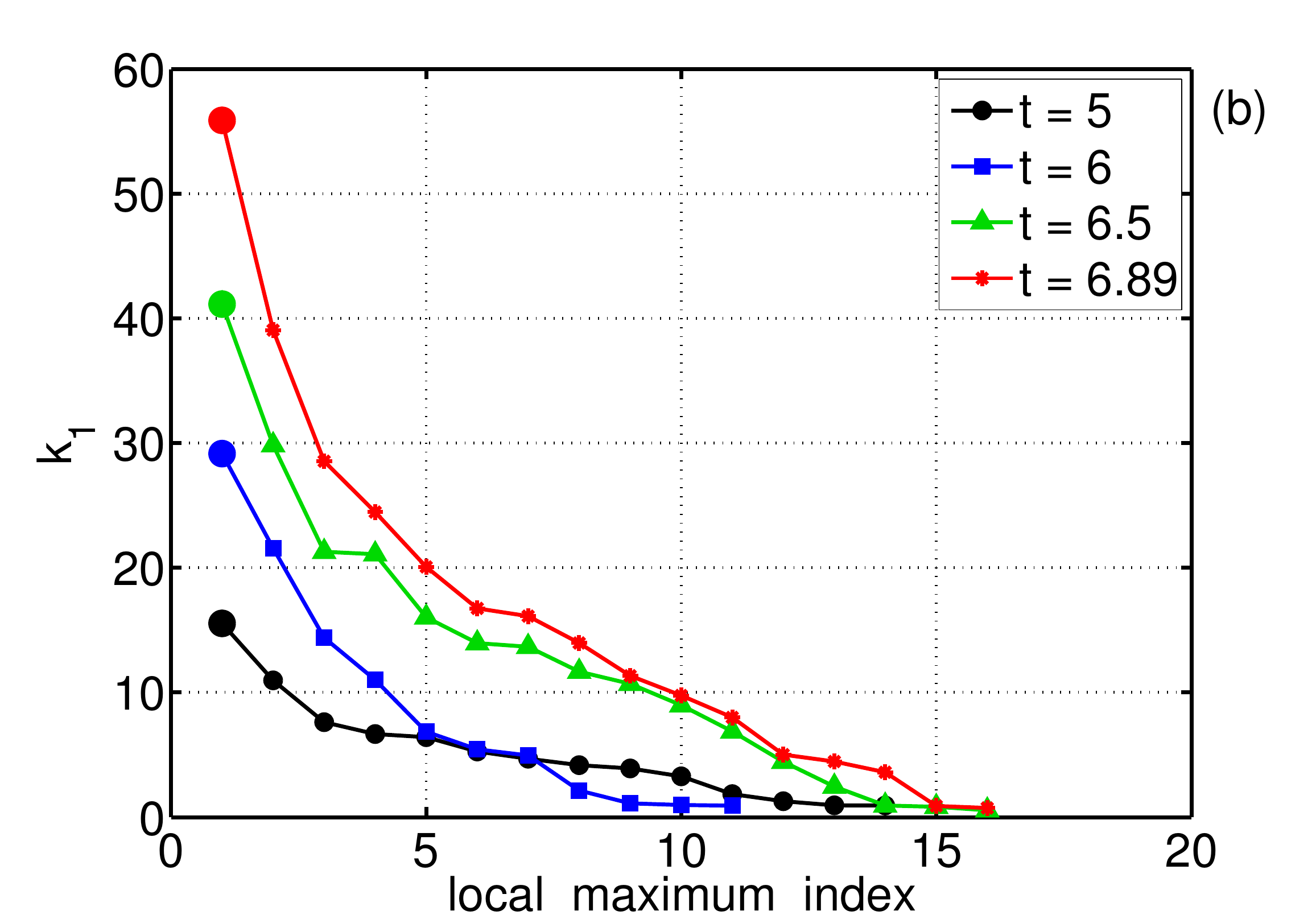}
\caption{{\it (Color on-line)} 
(a) Evolution of characteristic spatial scales $\ell_{1}$ (black), $\ell_{2}$ (blue) and $\ell_{3}$ (red) of local maximums, see Eq.~(\ref{Taylor_A123}). Exponentially decreasing values of $\ell_1$ indicate formation of the pancake structures in the vorticity field.
(b) Characteristic wavenumbers $k_{1}$ estimated by Eq.~(\ref{eq3}) for the local maximums versus local maximum index numbers at different times. Local maximums are sorted in decreasing order of $k_{1}$. Large circles mark the global maximums.}
\label{fig:evolution_lmB}
\end{figure}
	
Figure~\ref{fig:evolution_lmB}(a) shows the evolution of the three length scales, $\ell_1$, $\ell_2$ and $\ell_3$, computed for each local maximum 
according to Eq.~(\ref{Taylor_A123}). We see that the smallest scale $\ell_1$ (black) decays 
nearly exponentially, $\ell_1 \propto e^{-t/T_\ell}$, with rather close values of the characteristic time $T_{\ell}$ for different local maximums, while the scales $\ell_2$ (blue) and 
$\ell_3$ (red) remain the same or decrease slightly; 
recall that the large numerical error for the computation of $\ell_3$ is related to the ill-conditioned Hessian matrix. 
This figure confirms the visual observation (Fig.~\ref{fig5}) that most of the regions of increased vorticity tend 
to have the pancake shape characterized by the small spatial scale (thickness) $\ell_{1}$ exponentially decreasing with time.

It is very instructive to see the distribution of local maximums among the spatial scales at fixed time, which 
highlights the contribution of different pancake structures to the energy at different scales. 
Using the estimate (\ref{eq3}), we
plot in Fig.~\ref{fig:evolution_lmB}(b) the characteristic wavenumbers $k_{1}$ for local maximums in decreasing order. The results suggest that the two leading 
maximums propagate much faster towards large wavenumbers. However, all the other local maximums (i.e., most of the pancake structures) fill densely the interval from large to medium scales. Throughout this interval the distribution of maximums 
has a well-established slope (i.e., the pancakes are distributed with a specific ``spectral density''). This interval increases with time and reaches $0 < k_{1} \lesssim 30$ at the final time $t = 6.89$.

%----------------------------------------

\section{Vorticity structures and the Kolmogorov spectrum}
\label{SecKolm}

In fully developed turbulent flow with large Reynolds numbers, the energy from large scales is transported through a wide range of medium scales (the inertial interval) to small scales, where it is eventually dissipated, 
see~\cite{kolmogorov1941local,obukhov1941spectral,landau2013fluid,frisch1999turbulence}. The viscosity is negligible in the inertial interval, where the evolution can be described by the Euler equations. The dimensional considerations suggest the well-known Kolmogorov scaling law, $E_k \propto k^{-5/3}$, for the energy spectrum in the inertial interval. 
Our simulations can be seen as describing the initial stage of turbulent dynamics, 
developing from large-scale initial data, at times before the flow 
gets excited at viscous scales. 
Generally speaking, considerations of the Kolmogorov theory do not extend to this case, because 
the energy cascade is not formed yet. Thus, the scaling law $E_k \propto k^{-5/3}$ is not necessarily satisfied. 
From this point of view, our simulations of the 3D Euler equations give the possibility for studying the initial stage in formation of the Kolmogorov spectrum together with the respective energy transfer mechanisms.

\begin{figure}[t]
\centering
\includegraphics[width=8cm]{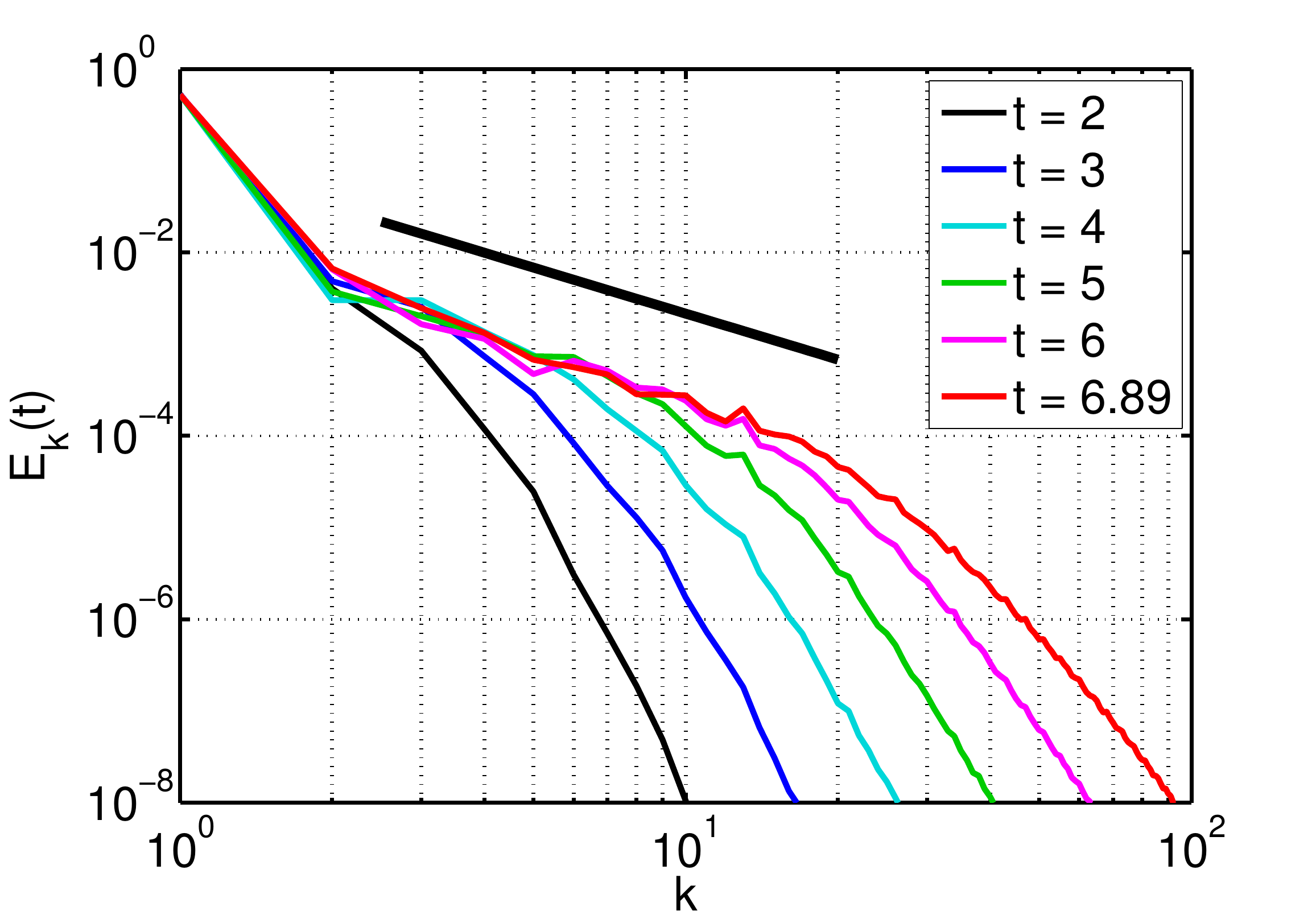}
\caption{{\it (Color on-line)} Energy spectrum $E_{k}(t)$ at different times. Straight line above the curves indicates the slope of the Kolmogorov power-law, $E_{k}\propto k^{-5/3}$.}
\label{fig:Ek_Sulem}
\end{figure}

Figure~\ref{fig:Ek_Sulem} shows that, at sufficiently small wavenumbers, we clearly observe the gradual formation of  the Kolmogorov interval $E_k \propto k^{-5/3}$. This interval grows with time and extends to a decade of wavenumbers, $2 \lesssim k \lesssim 20$, at the end of the simulation. The Kolmogorov interval corresponds to the ``frozen'' part of the energy spectrum: $E_{k}(t)$ changes slightly with time in the Kolmogorov region in contrast to the vast changes at larger wavenumbers. Taking into account the times and logarithmic scale in Figure~\ref{fig:Ek_Sulem}, one can guess that the size of the Kolmogorov interval increases exponentially in time. 

\begin{figure}[t]
\centering
\includegraphics[width=8cm]{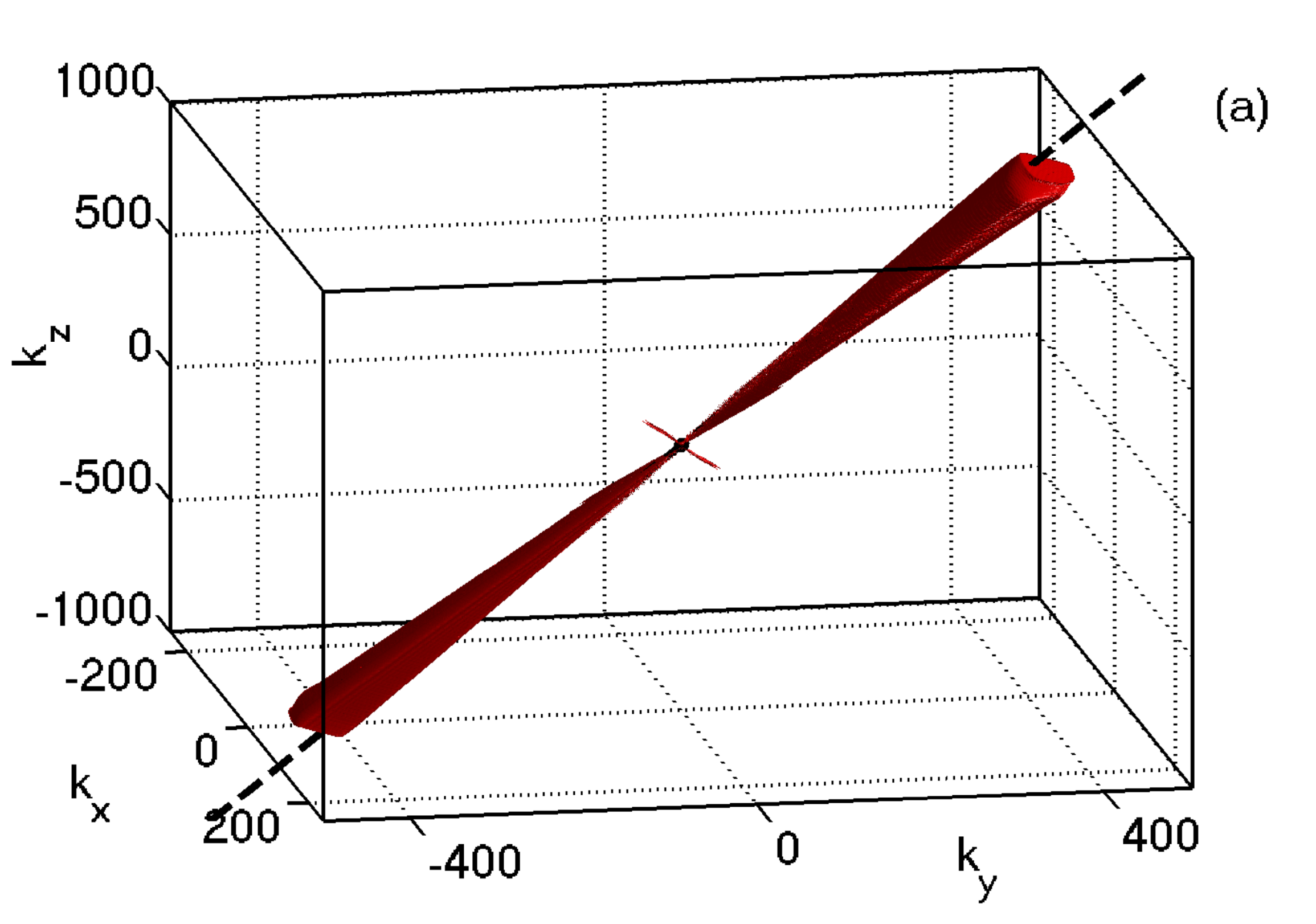}
\includegraphics[width=8cm]{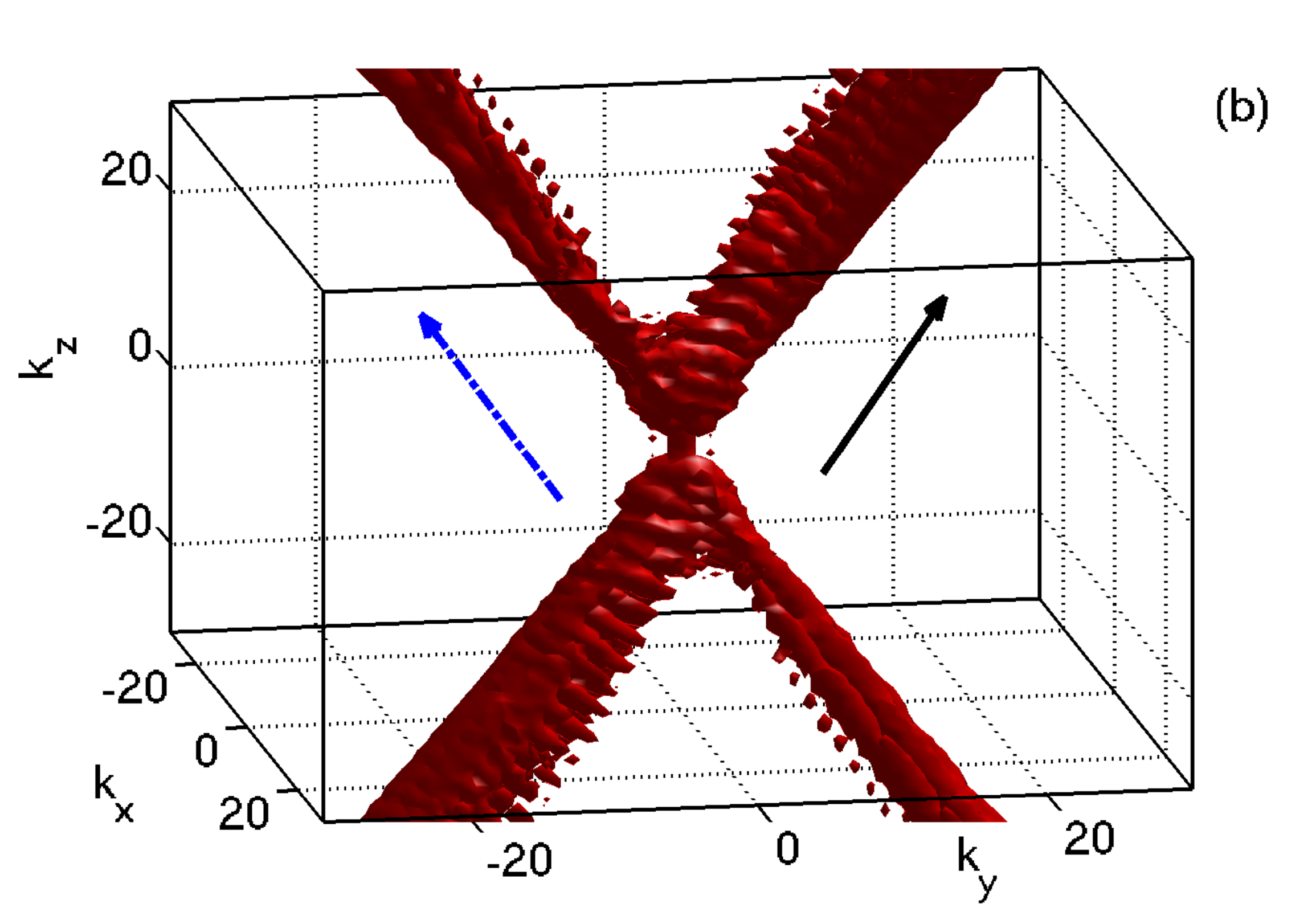}
\caption{{\it (Color on-line)} Isosurface $|\tilde{\omegabold}(\mathbf{k})| = 0.2$ 
of the normalized vorticity field (\ref{vorticity_renormalized}) in Fourier space at final time $t=6.89$. (a) The jet is aligned with the eigenvector $\mathbf{w}_{1}$ (dashed black line), which is the normal direction of the pancake structure at the global vorticity maximum in physical space. (b) The closer view with the eigenvector $\mathbf{w}_{1}$ (solid black arrow) 
for the global vorticity maximum and the respective eigenvector (dash-dot blue arrow) for the third largest local maximum.}
\label{fig:jet}
\end{figure}

\begin{figure}[t]
\centering
\includegraphics[width=8cm]{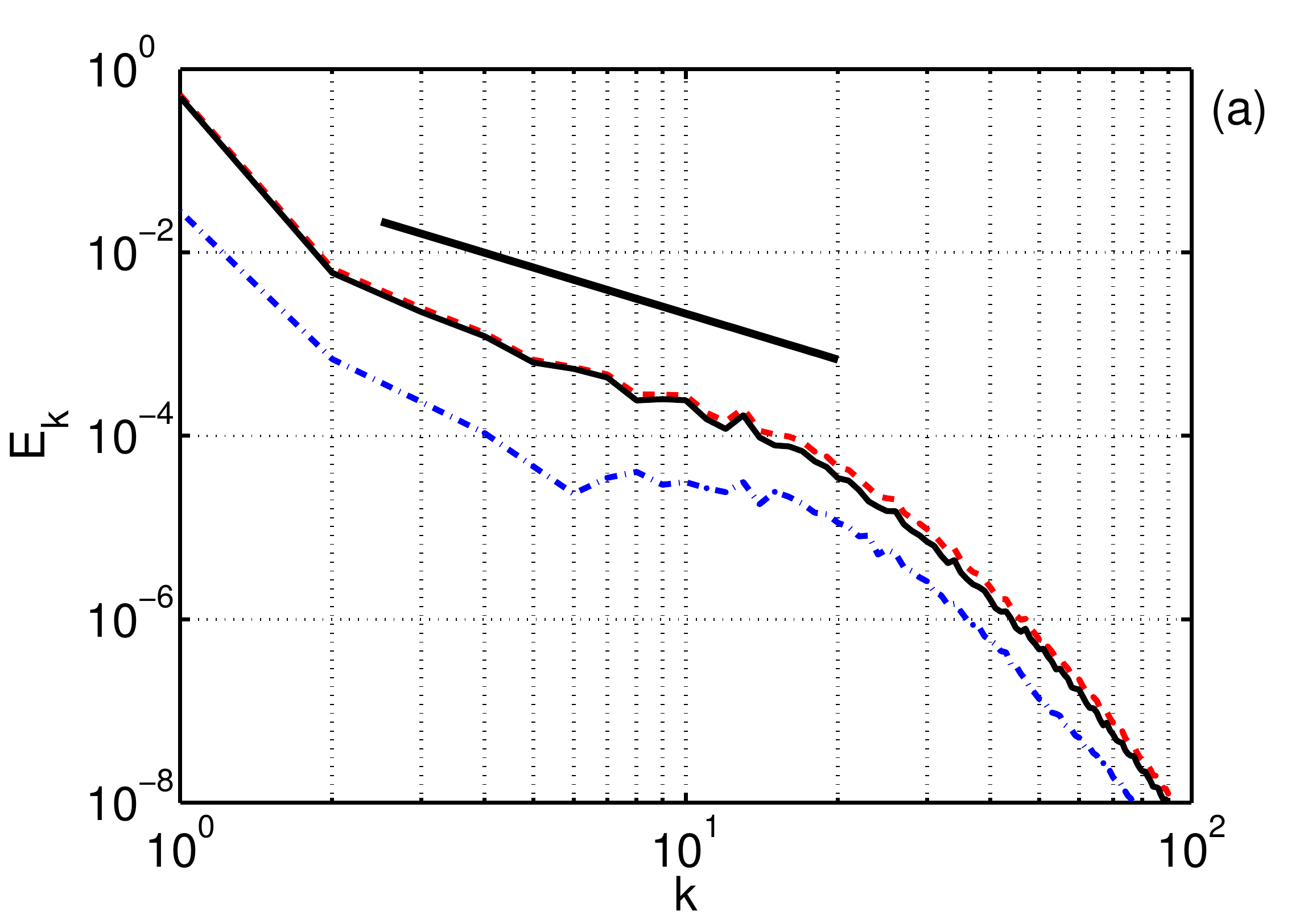}
\includegraphics[width=8cm]{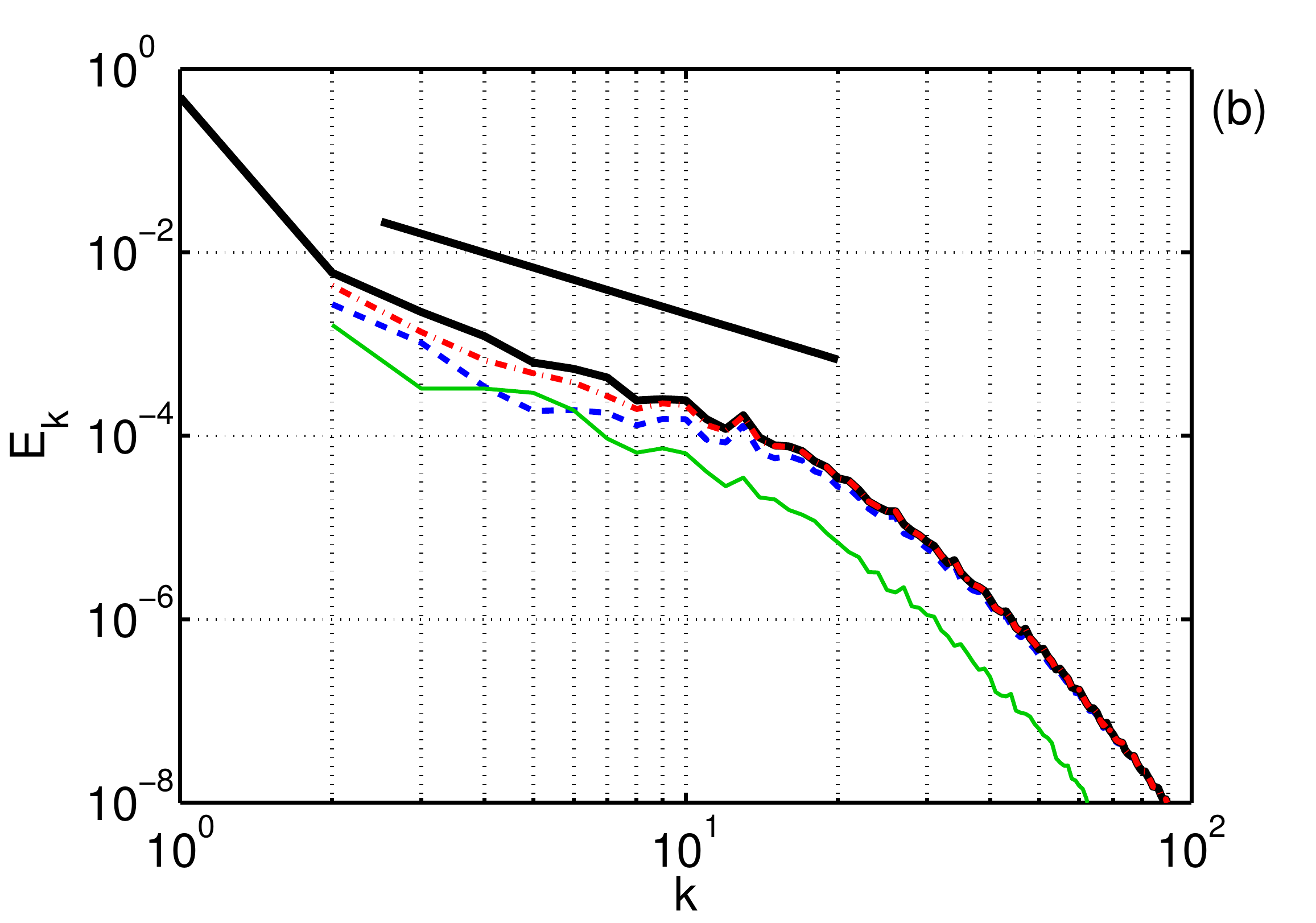}
\caption{{\it (Color on-line)} Energy spectrum $E_k$ at the end of the simulation. (a) Calculated inside the isosurface $|\tilde{\omegabold}(\mathbf{k})| > 0.2$ (solid black), outside this isosurface (dash-dot blue) and the sum of the two (dashed red).
(b) Calculated inside the isosurface $|\tilde{\omegabold}(\mathbf{k})| > 0.2$ (thick solid black), inside the first jet (dashed blue), inside the second jet (thin solid green), and the sum of the two jets (dash-dot red).
Black line above the curves indicates the slope of the Kolmogorov power-law, $E_{k}\propto k^{-5/3}$.}
\label{fig:jetB}
\end{figure}

In order to link the energy spectrum $E_k$ with the pancake vorticity structures studied in the previous Sections, 
we look at the flow in Fourier space.
One can expect that each pancake generates a structure extended in one direction in Fourier space (``jet''), 
aligned with the eigenvector $\mathbf{w}_{1}$ perpendicular to the pancake (such jets form, e.g.,  
in the two-dimensional inviscid flow~\cite{kuznetsov2007effects,kudryavtsev2012statistical}).   
Inside such a jet, the Fourier components of the flow should be large in comparison with the remaining background. In order to visualize this effect, we consider the function
\begin{equation}\label{vorticity_renormalized}
\tilde{\omegabold}(\mathbf{k}) = \omegabold(\mathbf{k})\Big/
\max_{|\mathbf{p}| = k}|\omegabold(\mathbf{p})|,
\end{equation}
representing the Fourier transformed vorticity $\omegabold(\mathbf{k})$ scaled to the maximal norm within each 
spherical shell. The reason for such a normalization is to compensate the strong decay of vorticity with $k$. 
Numerically, the maximum in Eq.~(\ref{vorticity_renormalized}) is computed among the nodes in a spherical shell of unit 
thickness. 

Figure~\ref{fig:jet}(a) shows the isosurface $|\tilde{\omegabold}(\mathbf{k})| = 0.2$, where the very thin interior part 
corresponds to larger vorticity, $|\tilde{\omegabold}(\mathbf{k})| > 0.2$. 
As expected, this isosurface is aligned with the eigenvector $\mathbf{w}_1$ computed at the global maximum, which should bring the dominant contribution to Fourier components of vorticity at large $k$. 
Figure~\ref{fig:jet}(b) demonstrates that the isosurface geometry is different at smaller wavenumbers, $k \lesssim 20$, corresponding to the Kolmogorov interval in Fig.~\ref{fig:Ek_Sulem}. 
Here, different jets contribute and 
some of these jets can be clearly related to the pancakes of other local vorticity maximums (the figure 
shows directions for the first and third largest local maximums, while 
the second local maximum yields the direction very close to that for the first one). 
As shown in Fig.~\ref{fig:jetB}(a), 
the small interior part of this isosurface dominates in the energy spectrum.
These observations reflect the extreme anisotropy of vorticity field and suggest that 
the Kolmogorov energy spectrum for $k \lesssim 20$  may be related to a collection of the pancake structures, rather than being determined solely by the dominant one. To further test this supposition, we integrate the energy spectrum inside each of the two jets $|\tilde{\omegabold}(\mathbf{k})| > 0.2$ shown in Fig.~\ref{fig:jet}(b), which can be 
separated from each other by cones of angle $\phi/2$, drawn from the origin around the corresponding eigenvectors $\mathbf{w}_1$. Here $\phi\approx\pi/4$ is the angle between the directions of the jets. One can see 
from Fig.~\ref{fig:jetB}(b) that the contributions of the jets to the energy spectrum are comparable at sufficiently small wavenumbers in the Kolmogorov region, while at larger wavenumbers (both inside and outside the Kolmogorov region) the contribution from the leading jet becomes dominant. Thus, the energy spectrum at sufficiently large wavenumbers $k\gtrsim 20$ is determined mainly by the leading pancake structure corresponding to the global vorticity maximum. This means that the process of energy transfer to small scales is performed through the evolution of the pancake vorticity structures in the physical space. This interpretation is similar, in a spirit, to the simplified turbulence models \cite{pullin1998vortex} based, e.g., on Lundgren vortices \cite{lundgren1982strained, gilbert1993cascade}. Note, however, that our simulation describes the initial stage of turbulence formation, when basic assumptions of these models related to stationary developed turbulence do not apply.

\begin{figure}[t]
\centering
\includegraphics[width=8cm]{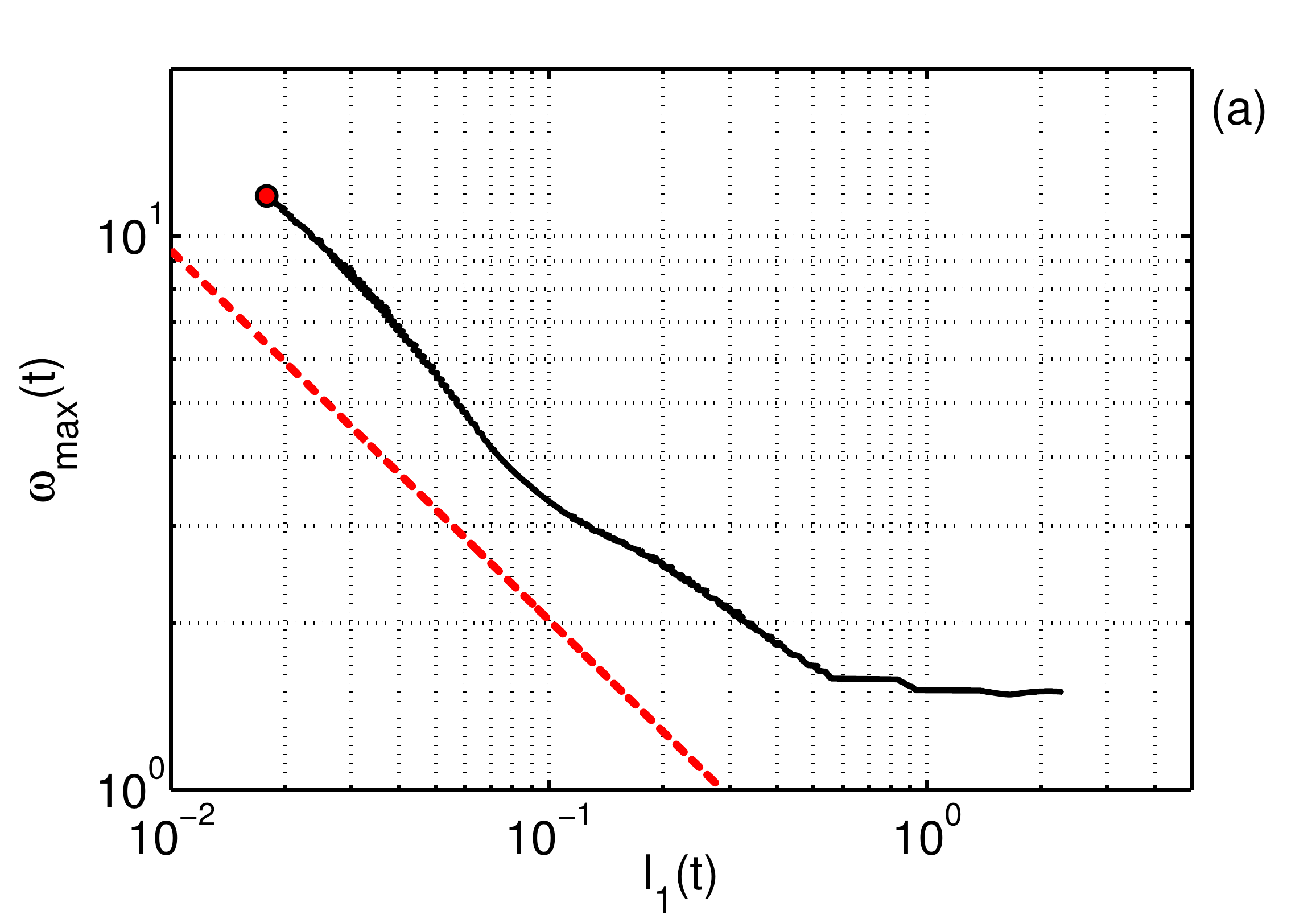}
\includegraphics[width=8cm]{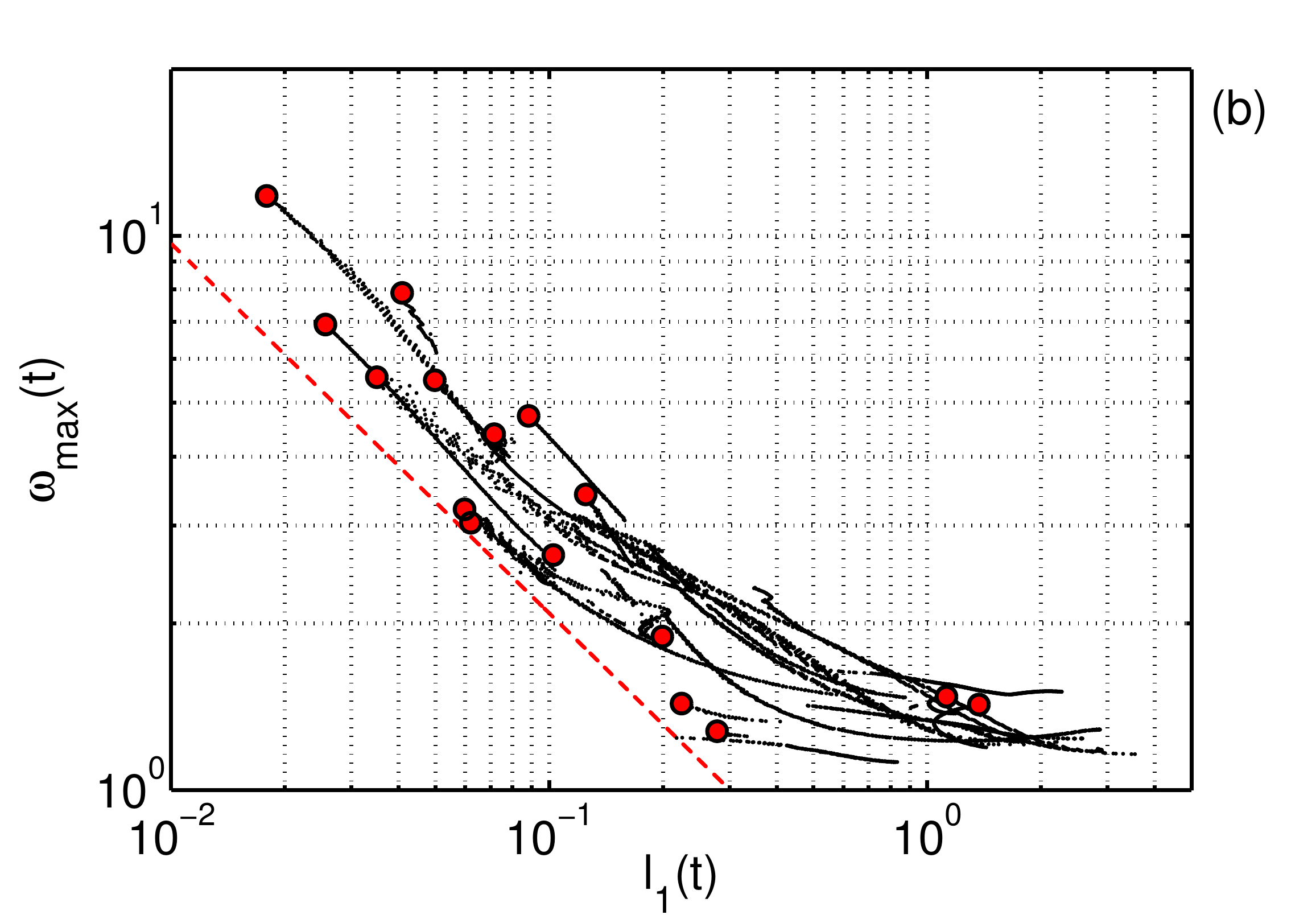}
\caption{{\it (Color on-line)} Relation between the vorticity local maximums $\omega_{\max}(t)$ and the respective characteristic lengths $\ell_{1}(t)$ during the evolution of the pancake structures: (a) for the global maximum, (b) for all local maximums. Red circles mark the local maximums at the final time $t=6.89$. Dashed lines indicate the power-law $\omega_{\max} \propto \ell_1^{-2/3}$ in Eq.~(\ref{singularity_w_to_dk}).}
\label{fig:W_vs_dk}
\end{figure}

Statistical properties of the pancakes analyzed in the previous Section can be elaborated further in relation to Kolmogorov theory. Notice that the Kolmogorov interval in Fig.~\ref{fig:Ek_Sulem} approximately coincides with the interval of wavenumbers in Fig.~\ref{fig:evolution_lmB}(b) occupied by most of the pancake structures (except for the few largest local maximums with considerably larger wavenumbers). The Kolmogorov theory suggests the power-law $\langle|\delta v|\rangle \propto \ell^{1/3}$ for the mean velocity variation at spatial scales $\ell$ in the inertial interval, see e.g.~\cite{landau2013fluid,frisch1999turbulence}. Similarly, it implies $\langle|\delta\omega|\rangle \propto \ell^{-2/3}$ for the vorticity variation. One can guess that the last relation is mainly determined by the small-scale high vorticity regions. In our simulations, that can be seen as describing the initial stage of turbulent dynamics, these high vorticity regions appear in the form of the pancake 
structures. As we argued in the previous Section, each pancake is characterized by the spatial scale $\ell_1(t) \propto e^{-t/T_{\ell}}$ (the other two scales $\ell_2$ and $\ell_3$ remain close to $1$) and the vorticity variation near the pancake is $|\delta\omega| \sim \omega_{\max}(t) \propto e^{t/T_\omega}$. Thus, we can test whether relation 
\begin{equation}\label{singularity_w_to_dk}
\omega_{\max}(t) \propto \ell_{1}(t)^{-2/3},
\end{equation}
holds during the evolution of the pancake structures. 

As shown in Fig.~\ref{fig:W_vs_dk}(a), the global vorticity maximum indeed evolves along the $\ell_{1}^{-2/3}$ law. This can be checked additionally by the ratio of the characteristic times $T_{\ell}$ and $T_{\omega}$, that, according to relation (\ref{singularity_w_to_dk}), should be equal to $T_{\ell}/T_{\omega}=2/3$. Our simulation provides the close relation $T_{\ell}/T_{\omega} \approx 0.7$ for the global vorticity maximum. It is also close for the simulation of the second initial condition $I_2$, which yields $T_{\ell}/T_{\omega}\approx 0.63$, see Appendix~\ref{App:B}. As one can see from Fig.~\ref{fig:W_vs_dk}(b), where the values of $\omega_{\max}$ are plotted versus the characteristic lengths $\ell_1$ for all the local maximums, most local maximums have the tendency to follow the power law (\ref{singularity_w_to_dk}) asymptotically. Note that the observed behavior agrees with the asymptotic pancake dynamics in accordance with Eqs.~(\ref{eq2})-(\ref{eq2_1}). However, this model allows 
any value for the ratio 
$T_{\ell}/T_{\omega}$. The blowup scenario 
based on the vortex lines breaking~\cite{kuznetsov2000collapse, kuznetsov2004breaking} leads to the $2/3$ ratio for the pancake structure $T_{\ell}/T_{\omega}$ and, thus, it can be considered as a possible theoretical justification if extended to the case of the exponential vorticity growth.
It is remarkable that local maximums follow approximately the same scaling law (\ref{singularity_w_to_dk}), with some shared constant prefactor before $\ell_{1}^{-2/3}$, and also that all local maximums taken at fixed time are distributed around this law, see red points in Fig.~\ref{fig:W_vs_dk}(b). These two properties indicate strong correlation of pancakes in the process of energy transport to small scales at initial stages of turbulent flow. 

%----------------------------------------
%----------------------------------------

\section{Conclusions}
\label{SecConcl}

In this work we performed the systematic numerical study of high vorticity structures that develop in the 3D incompressible Euler equations from smooth initial conditions of finite energy. Being motivated by the open problem of the finite-time blowup and its 
role for the developed turbulence, we are led to two important observations. First, we show that the exponential growth of vorticity, which is typically observed within pancakes (thin and wide vortex sheets), 
is compatible with the formation of Kolmogorov turbulent spectra in the fully inviscid flow, i.e., 
before the viscous scales get excited. Second, we show that the pancake structures, which have self-similar dynamics 
and develop in increasing number, play the crucial role in formation of the energy cascade to small scales. 

We demonstrated that the thickness of pancake-like regions of high vorticity decreases exponentially in time, 
$\ell_{1}(t)\propto e^{-t/T_{\ell}}$, while the other two dimensions do not change considerably, $\ell_{2}\sim\ell_{3}\sim 1$. 
At the same time the local vorticity maximum grows exponentially, $\omega_{\max}(t)\propto e^{t/T_{\omega}}$. 
During the evolution, the relation $\omega_{\max}(t)\propto \ell_{1}(t)^{-2/3}$ resembling the Kolmogorov scaling law holds 
approximately in agreement with the ratio $T_{\ell}/T_{\omega} \approx 2/3$ between the characteristic times of the pancake compression $T_{\ell}$ and the vorticity growth $T_{\omega}$.
Since the pancake 
evolution is governed by the two different time scales $T_{\ell}$ and $T_{\omega}$, this behavior does 
not agree with the pancake model (\ref{eq1}) proposed in 
\cite{brachet1992numerical}. However, the modified model (\ref{eq2})-(\ref{eq2_1}) satisfies the Euler equations for the leading terms and adequately describes the observed dynamics. The total number of pancake structures, estimated by the number of local vorticity maximums, increases with time. We demonstrate that at late times most of the pancakes are distributed 
densely across the corresponding interval of wavenumbers. 

We clearly observe the formation of the Kolmogorov energy spectrum $E_{k}\propto k^{-5/3}$ in the inviscid system, 
together with the exponential (i.e., no finite-time blowup) vorticity growth. The interval with Kolmogorov scaling grows with time and extends to a decade of wavenumbers at the end of the simulations. The energy spectrum $E_{k}(t)$ changes weakly with time in this region in contrast to vast changes at larger wavenumbers. 

Thin pancake structures in physical space generate strongly anisotropic vorticity field in Fourier space in the form of ``jets'', which are extended in the directions perpendicular to the pancakes. Within these jets the Fourier components of the flow are large in comparison with the remaining background. We demonstrate that these jets occupy a small fraction of the entire spectral band, but provide the leading contribution to the energy spectrum of the system. This means that the energy transfer to small scales is performed through the evolution of the pancake structures, and that the Kolmogorov energy spectrum may be attributed to collective behavior of the pancakes. 

\begin{acknowledgements}
The authors thank M. Fedoruk for the access to and V. Kalyuzhny for the assistance with Novosibirsk Supercomputer Center. The work of D.S.A. and E.A.K was supported by the RSF (Grant No. 14-22-00174). D.S.A. is grateful to IMPA for the support of his visits to Brazil. The work of A.A.M was supported by the CNPq (Grant No. 305519/2012-3), Program FAPERJ Pensa Rio--2014, and RFBR (Grant No. 13-01-00261).
\end{acknowledgements}

%\pagebreak

\appendix
\section{Initial conditions}
\label{App:A}

\renewcommand{\arraystretch}{0.8}

\begin{table}[H]
\caption{Nonzero coefficients in Eq.~(\ref{IC1}) for the initial vorticity field $I_{1}$ with the average energy density $E/(2\pi)^{3}\approx 0.54$ and the average helicity density $\Omega/(2\pi)^{3}\approx 1.05$. Final time for this simulation is $t=6.89$ and the final grid is $486\times 1024\times 2048$. The simulation is affected by aliasing starting from $t\approx 5.13$. }
\begin{center}
 \begin{tabular}{| c | c | c |}
  \hline
  $\mathbf{h}$ & $\mathbf{A_{h}}$ & $\mathbf{B_{h}}$ \\ \hline
  (-1,0,2) & (0.0065641, 0.0027931, 0.003282) & (0.0044136, 0.0056271, 0.0022068) \\ \hline
  (0,0,0) & (0.065101, 0.0005801, -0.064109) & (0.0045744, -0.022895, 0.18392) \\ \hline
  (0,0,1) & (0, 1, 0) & (1, 0, 0) \\ \hline
  (0,0,2) & (0, 0.01, 0) & (0.01, 0, 0) \\ \hline
  (0,1,0) & (0.21204, 0, -0.070625) & (-0.14438, 0, 0.23298) \\ \hline
  (0,1,1) & (0.045977, -0.010151, 0.010151) & (0.041942, 0.040326, -0.040326) \\ \hline
  (0,2,0) & (0.005, 0, 0) & (0, 0, 0.005) \\ \hline
  (1,0,0) & (0, 0, 0.1) & (0, 0.1, 0) \\ \hline
  (1,0,1) & (-0.046112, 0.017081, 0.046112) & (-0.0097784, 0.020122, 0.0097784) \\ \hline
  (1,1,2) & (-0.0034664, 0.0049556, -0.00074462) & (-0.0059316, -0.0010472, 0.0034894) \\ \hline
  (2,0,0) & (0, 0, 0.02) & (0, 0.02, 0) \\ \hline
 \end{tabular}
\end{center}
 \label{tab:M1}
\end{table}

\begin{table}[H]\footnotesize
\caption{Nonzero coefficients in Eq.~(\ref{IC1}) for the initial vorticity field $I_{2}$ with the average energy density 
$E/(2\pi)^{3}\approx 0.51$ and the average helicity density $\Omega/(2\pi)^{3}\approx 1$. 
Final time for this simulation is $t=7.77$ and the final grid is $1152\times 384\times 2304$. The simulation is affected by aliasing starting from $t\approx 6.25$.}
\begin{center}
 \begin{tabular}{| c | c | c |}
  \hline
  $\mathbf{h}$ & $\mathbf{A_{h}}$ & $\mathbf{B_{h}}$ \\ \hline
  (-1,0,1) & (-0.040618, 0.039651, -0.040618) & (-0.030318, 0.064657, -0.030318) \\ \hline
  (0,0,0) & (0.067751, -0.1311, -0.11256) & (-0.082614, -0.0364, 0.18932) \\ \hline
  (0,0,1) & (0, 1, 0) & (1, 0, 0) \\ \hline
  (0,1,1) & (0.0062549, 0.044315, -0.044315) & (0.034983, -0.014521, 0.014521) \\ \hline
  (1,0,0) & (0, 0.079395, 0.07027) & (0, 0.099411, 0.012762) \\ \hline
  (1,1,0) & (-0.047174, 0.047174, -0.045572) & (-0.049622, 0.049622, 0.001773) \\ \hline
 \end{tabular}
\end{center}
 \label{tab:M2}
\end{table}

%----------------------------------------
%----------------------------------------

\section{Simulation results for the initial condition $I_{2}$}
\label{App:B}

\begin{figure}[H]
\centering
\includegraphics[width=8cm]{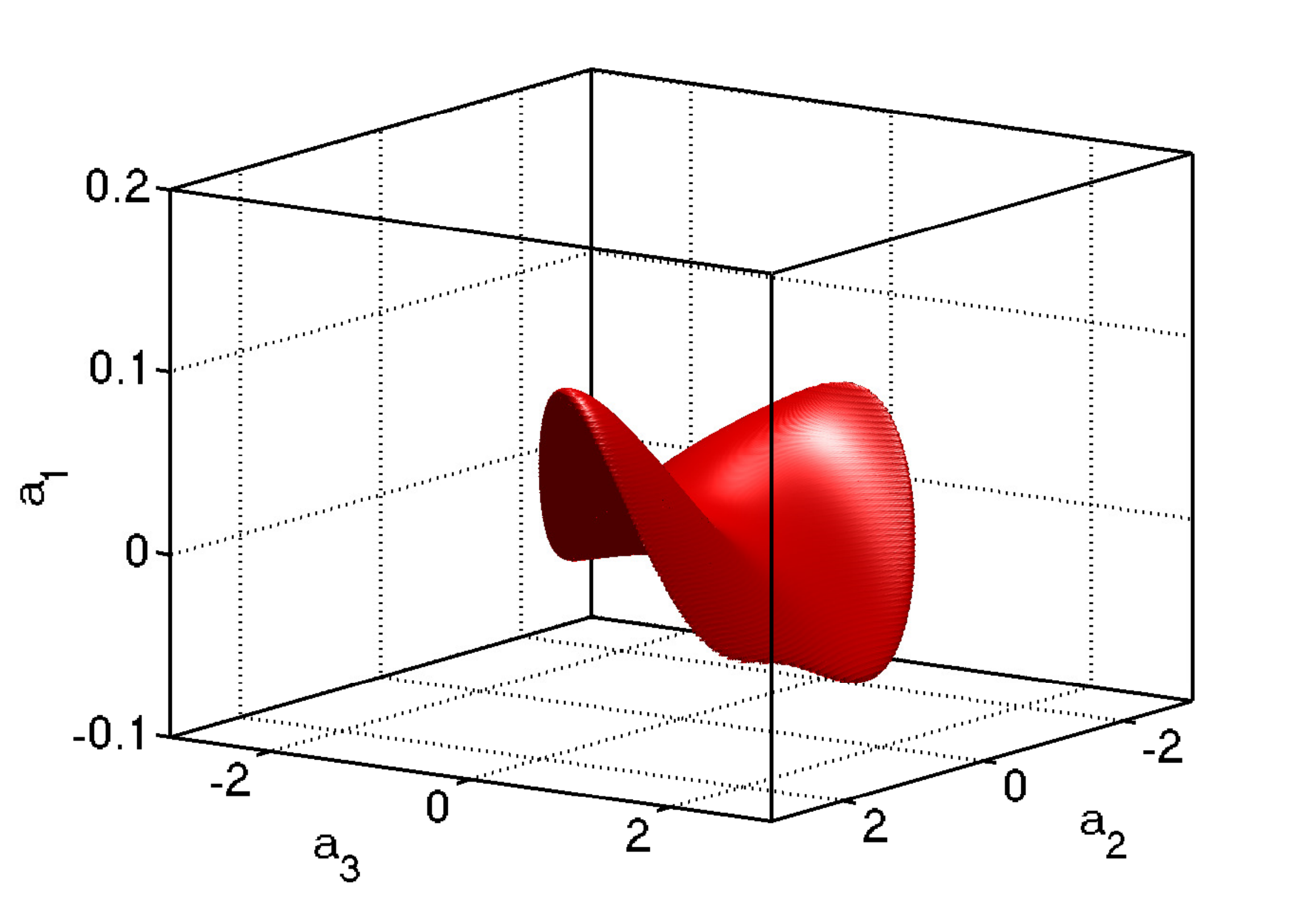}
\caption{{\it (Color on-line)} Isosurface of constant vorticity $|\omegabold|=0.8\,\omega_{\max}$ in the local coordinates $(a_{1},a_{2},a_{3})$ at the final time of the simulation, $t=7.77$. Note much smaller vertical scale.}
\label{fig:2nd_evolution_lm_dimensions_iso}
\end{figure}

\begin{figure}[H]
\centering
\includegraphics[width=8cm]{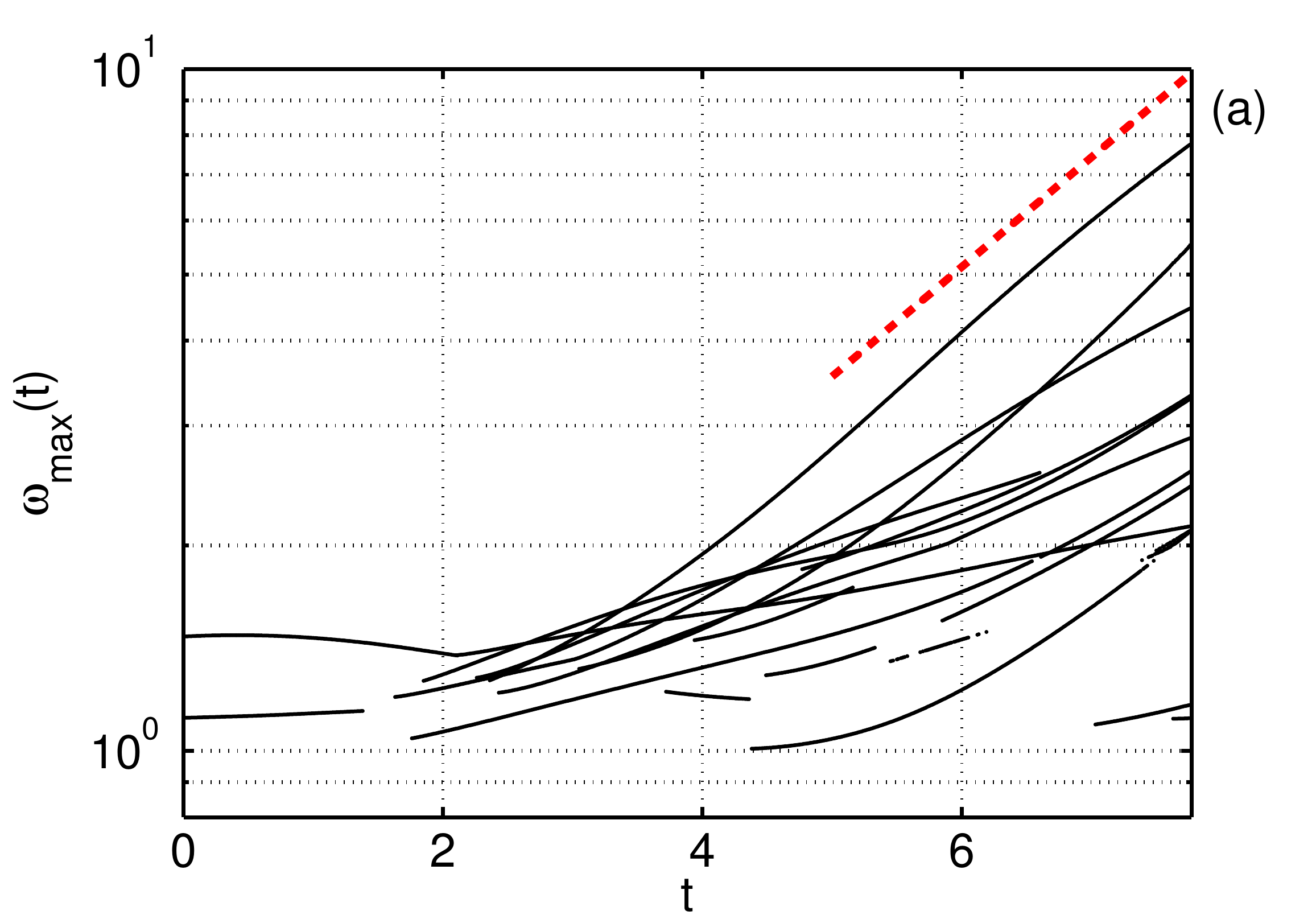}
\includegraphics[width=8cm]{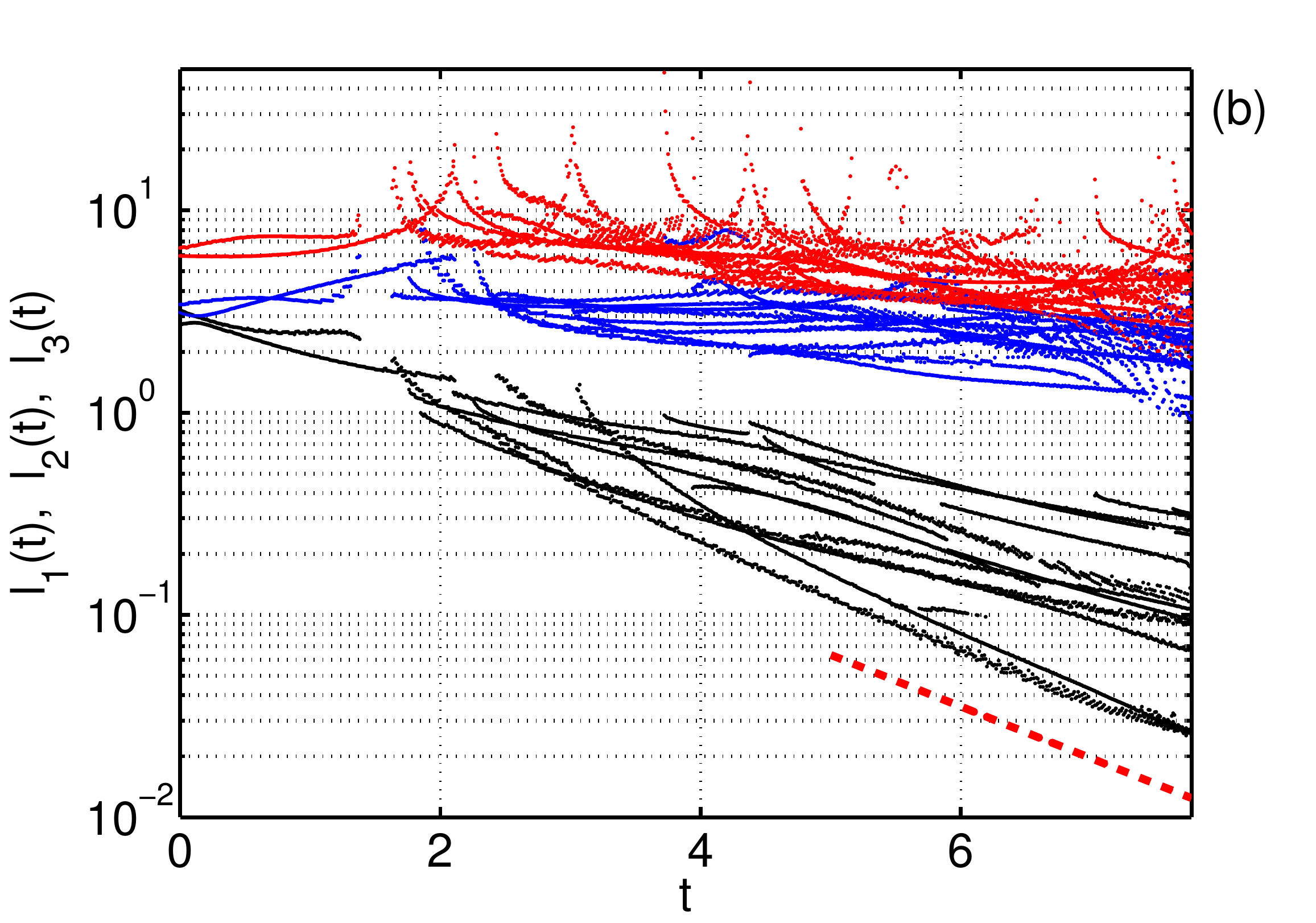}
\caption{{\it (Color on-line)} (a) Evolution of local vorticity maximums (logarithmic vertical scale). 
The dashed red line indicates the exponential slope $\propto\!e^{t/T_{\omega}}$ with characteristic time $T_{\omega} = 2.7$.
(b) Evolution of characteristic spatial scales $\ell_{1}$ (black), $\ell_{2}$ (blue) and $\ell_{3}$ (red) of local maximums, see Eq.~(\ref{Taylor_A123}). The dashed red line indicates the exponential slope $\propto\!e^{-t/T_{\ell}}$ with characteristic time $T_{\ell} = 1.7$. }
\label{fig:2nd_evolution_lm_dimensions}
\end{figure}

In this Appendix we provide some results for the simulation with initial condition $I_{2}$, which reaches the time $t=7.77$ on the grid $1152\times 384\times 2304$. The region of global vorticity maximum represents at final time a very thin pancake structure, as shown in Fig.~\ref{fig:2nd_evolution_lm_dimensions_iso} by the numerically computed isosurface $|\omegabold(\mathbf{r})| = 0.8\,\omega_{\max}$ in the local coordinates $(a_{1},a_{2},a_{3})$, see Eq. (\ref{Taylor_A123}). The number of local maximums increases with time, from 2 at $t=0$ to 13 at the end of the simulation. The values of vorticity modulus at maximum points tend to grow exponentially in time, $\omega_{\max}(t)\propto e^{t/T_{\omega}}$, with different but relatively close values of the characteristic times $T_{\omega}$, Fig.~\ref{fig:2nd_evolution_lm_dimensions}(a). The thickness of the associated high vorticity regions decays nearly exponentially, $\ell_{1}(t)\propto e^{-t/T_{\ell}}$, 
while the other two scales $\ell_{2}$ and $\ell_{3}$ remain the same or decrease slightly in time, Fig.~\ref{fig:2nd_evolution_lm_dimensions}(b). 

\begin{figure}[H]
\centering
\includegraphics[width=8cm]{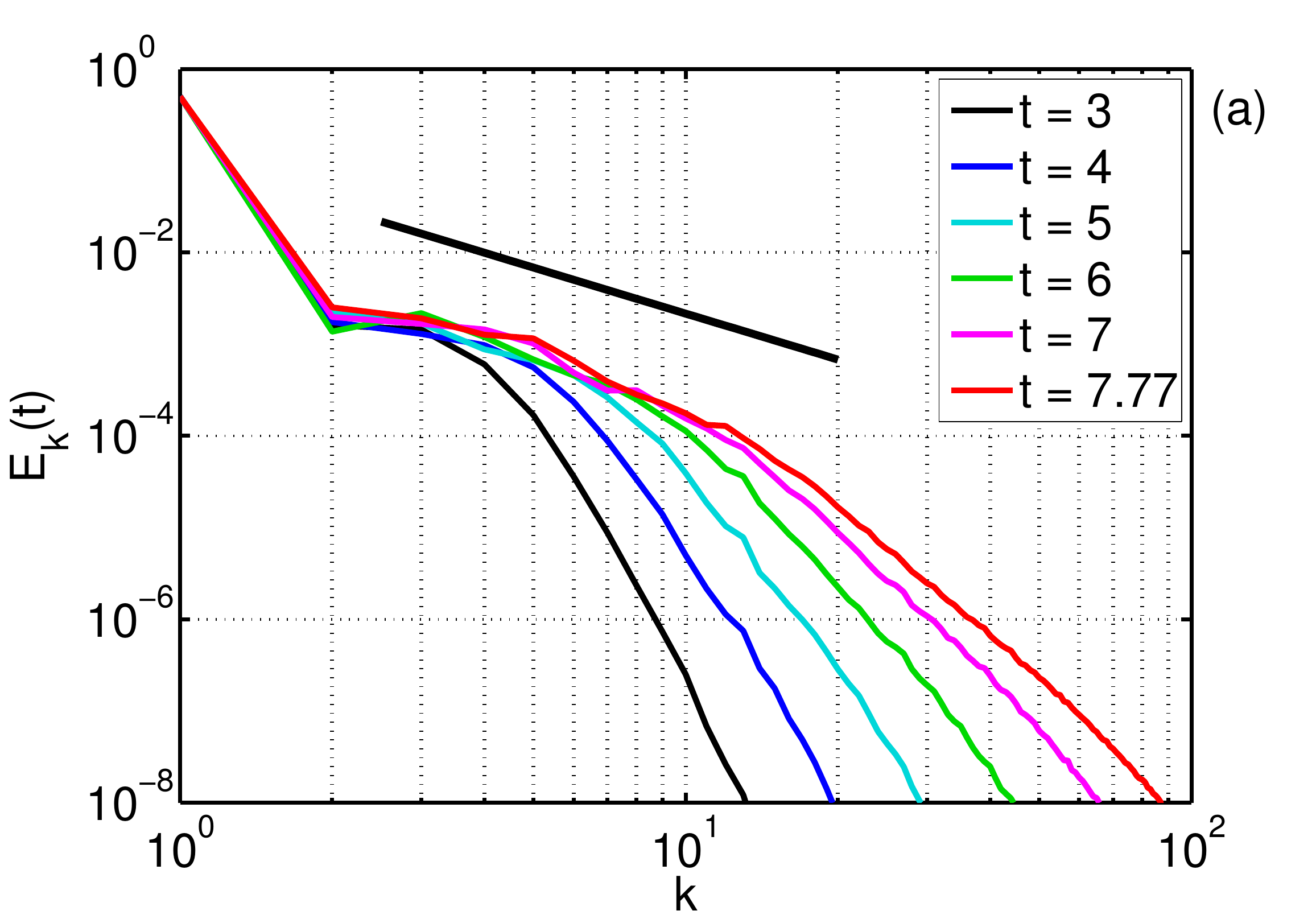}
\includegraphics[width=8cm]{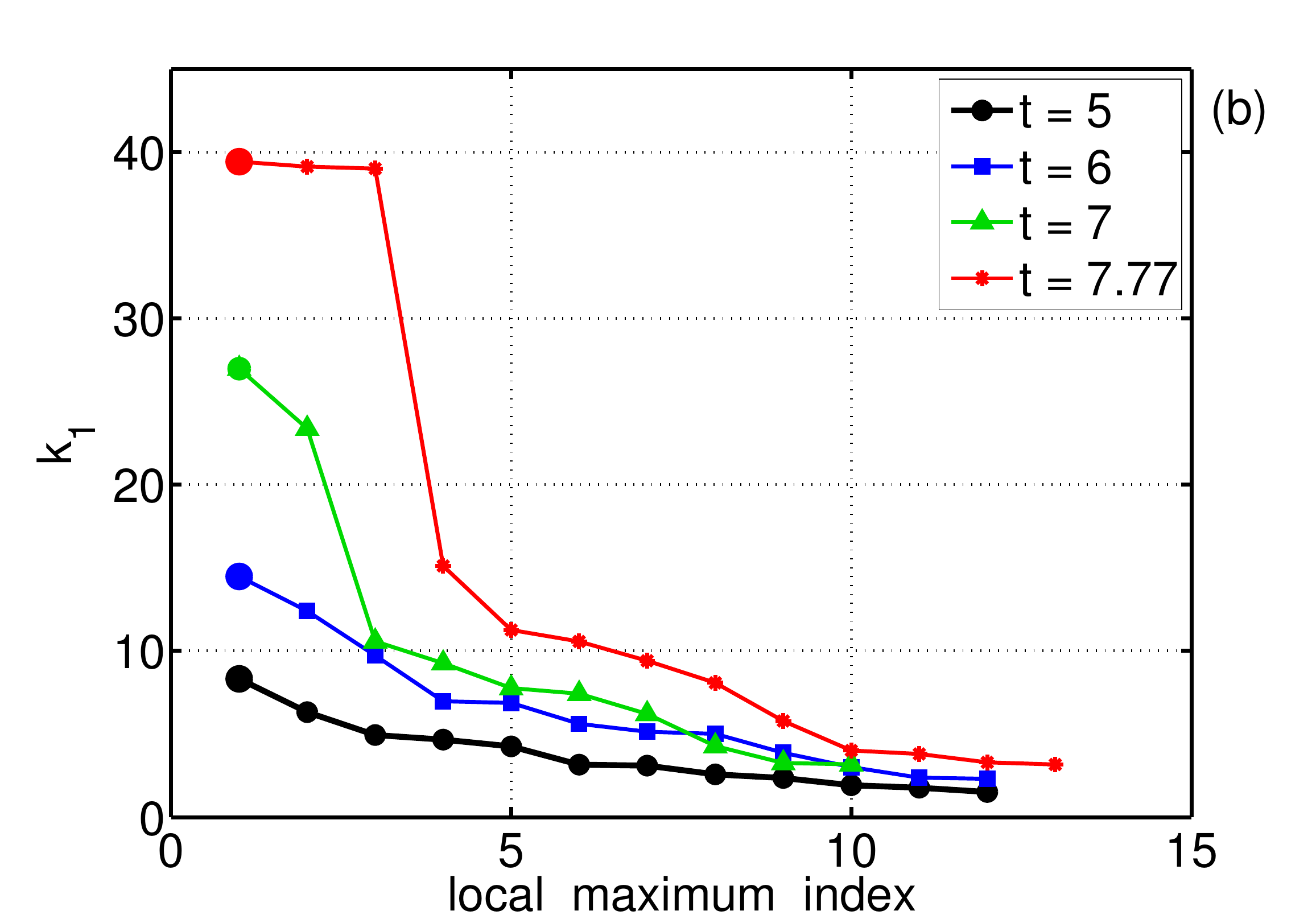}
\caption{{\it (Color on-line)} 
(a) Energy spectrum $E_{k}(t)$ at different times. Straight line above the curves indicates the slope of the Kolmogorov power-law, $E_{k}\propto k^{-5/3}$.
(b) Characteristic wavenumbers $k_{1}$ estimated by Eq.~(\ref{eq3}) for the local maximums versus local maximum index numbers at different times. Local maximums are sorted in decreasing order of $k_{1}$. Large circles mark the global maximums.}
\label{fig:2nd_evolution_lmB}
\end{figure}

The gradual formation of the Kolmogorov region $E_{k}\propto k^{-5/3}$ in the energy spectrum is shown in Fig.~\ref{fig:2nd_evolution_lmB}(a). This region extends to $5\lesssim k\lesssim 15$ at the end of the simulation, 
and corresponds to the ``frozen'' part of the spectrum where $E_{k}(t)$ changes slightly in time, in contrast to the vast changes at larger wavenumbers. 
The size of this interval is smaller than for the first simulation in Fig.~\ref{fig:Ek_Sulem}, but the same tendency is clearly observed.
Fig.~\ref{fig:2nd_evolution_lmB}(b) shows the characteristic wavenumbers $k_{1}=1/\ell_{1}$ for local maximums in decreasing order. The first three local maximums propagate much faster to higher wavenumbers, while the other local maximums fill densely the interval from large to medium scales, which approximately coincides with the Kolmogorov region in the energy spectrum.

\begin{figure}[H]
\centering
\includegraphics[width=8cm]{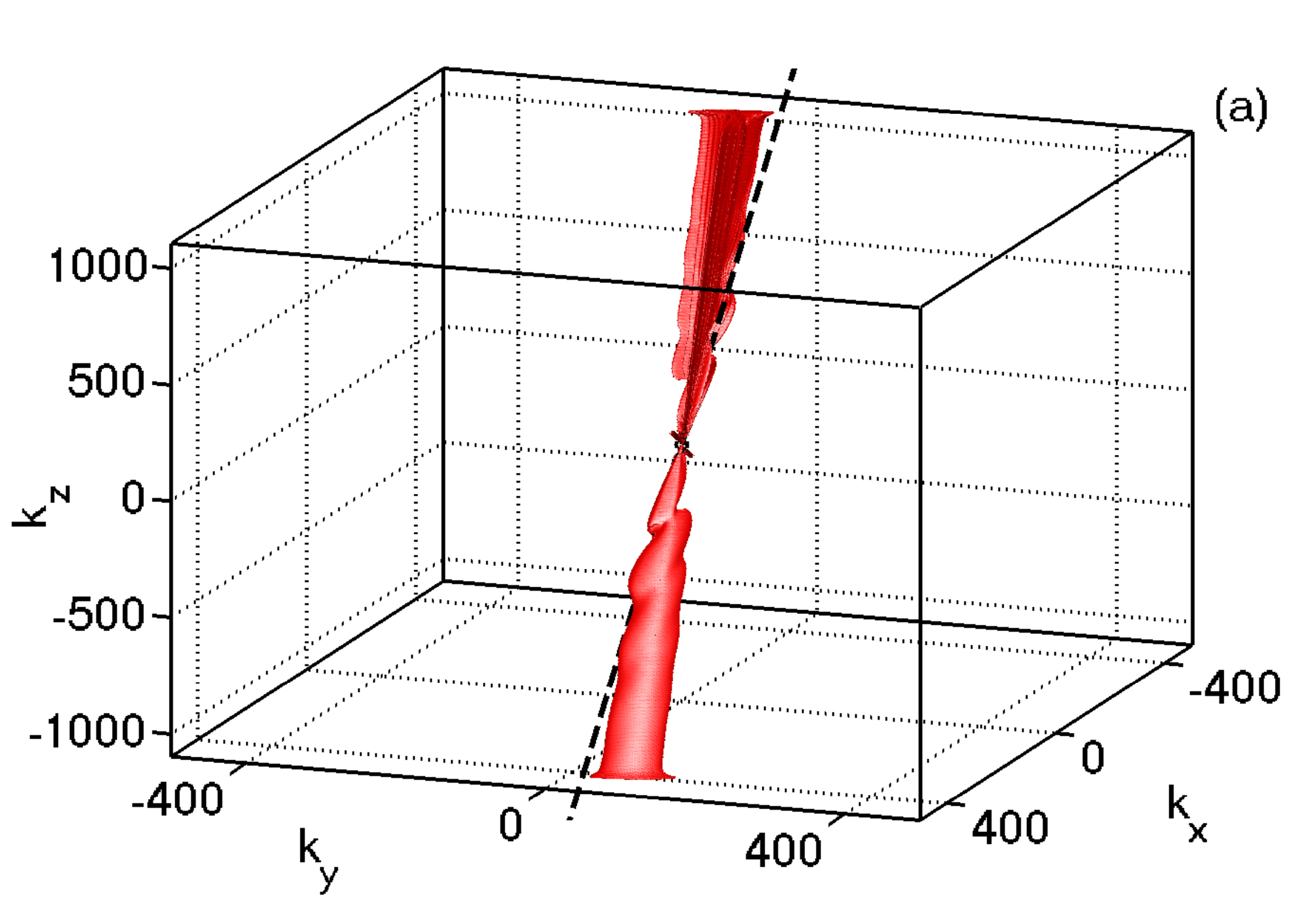}
\includegraphics[width=8cm]{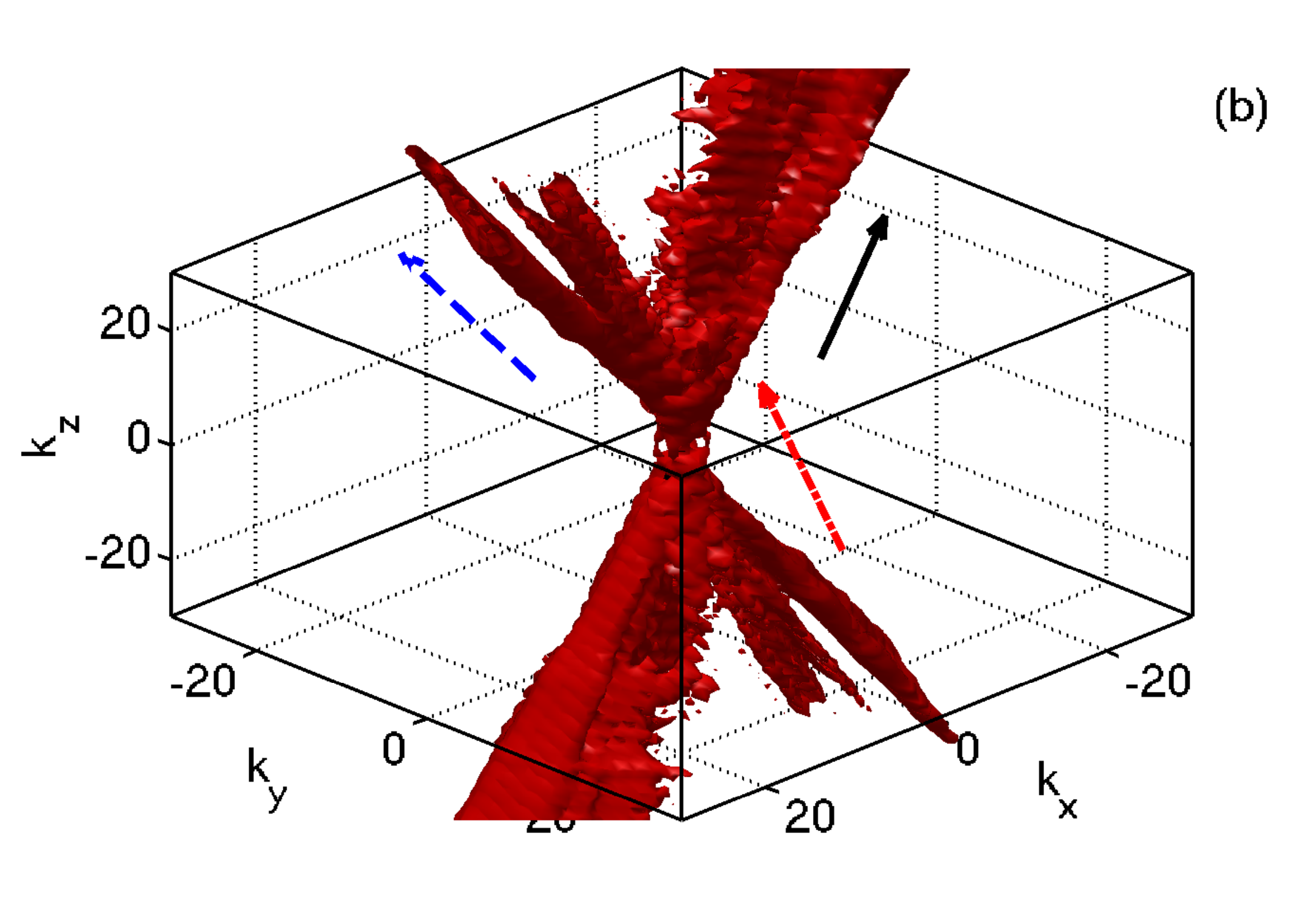}
\caption{{\it (Color on-line)} Isosurface $|\tilde{\omegabold}(\mathbf{k})| = 0.2$ 
of the normalized vorticity field (\ref{vorticity_renormalized}) in Fourier space at final time $t=7.77$. (a) The jet is aligned with the eigenvector $\mathbf{w}_{1}$ (dashed black line), which is the normal direction of the pancake structure at the global vorticity maximum in physical space. (b) The closer view with the eigenvector $\mathbf{w}_{1}$ (solid black arrow) 
for the global vorticity maximum and the similar eigenvectors for the second (dashed blue arrow) and the fourth (dash-dot red arrow) largest local maximums.}
\label{fig:2nd_jet}
\end{figure}

\begin{figure}[H]
\centering
\includegraphics[width=8cm]{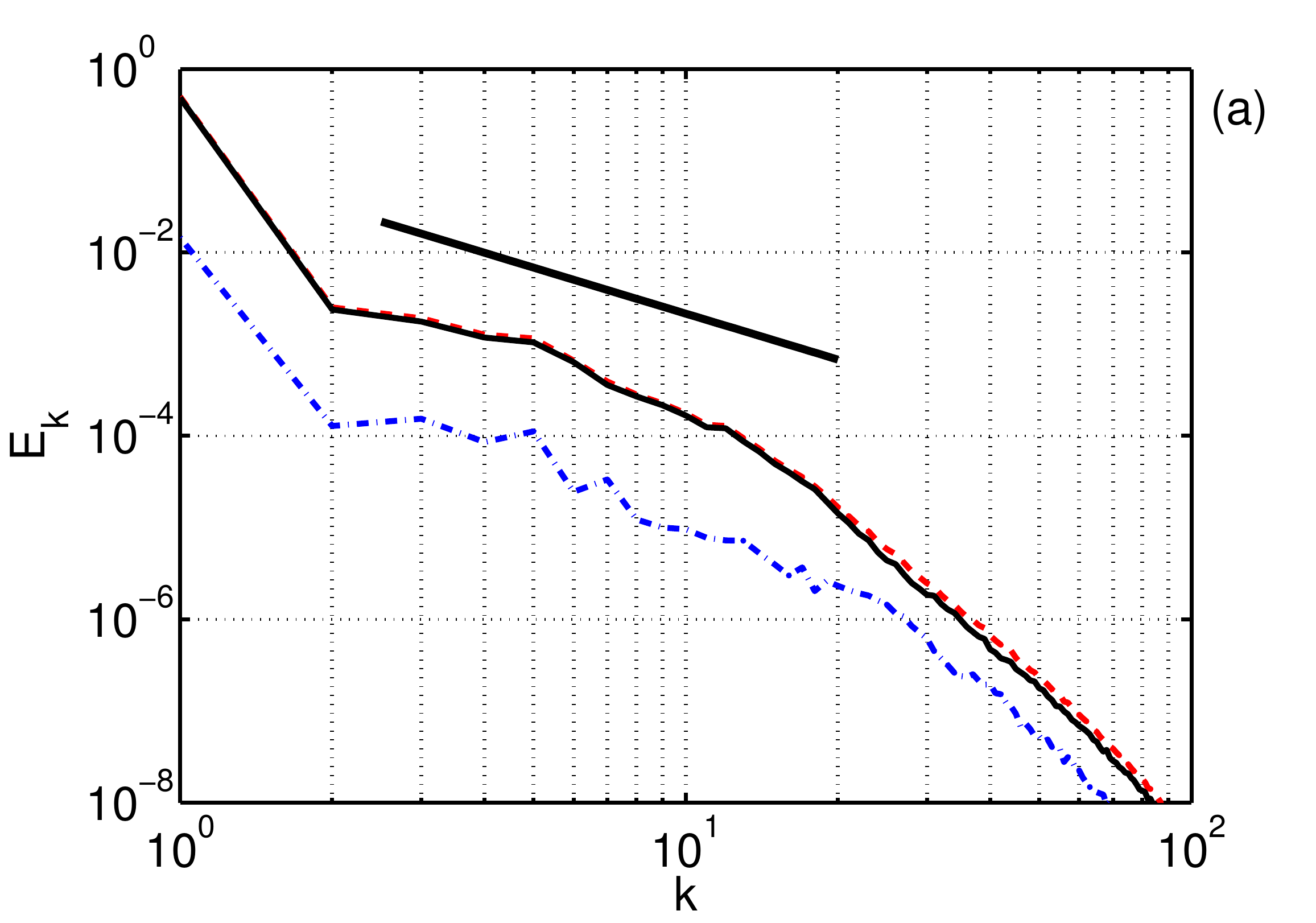}
\includegraphics[width=8cm]{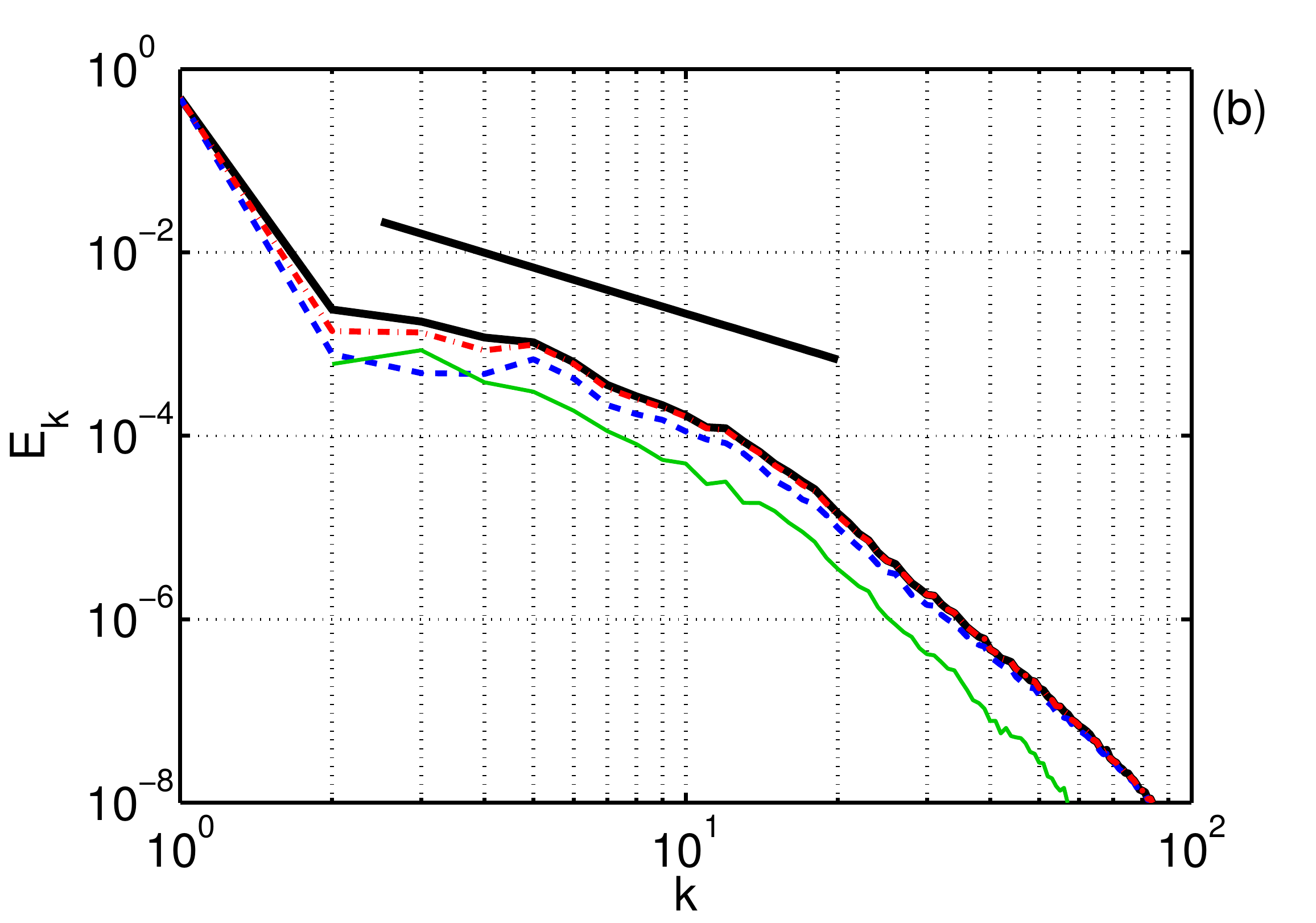}
\caption{{\it (Color on-line)} Energy spectrum $E_k$ at the end of the simulation. (a) Calculated inside the isosurface $|\tilde{\omegabold}(\mathbf{k})| > 0.2$ (solid black), outside this isosurface (dash-dot blue) and the sum of the two (dashed red).
(b) Calculated inside the isosurface $|\tilde{\omegabold}(\mathbf{k})| > 0.2$ (thick solid black), inside the first jet (dashed blue), inside the second and third jets (thin solid green), and the sum of the three jets (dash-dot red).
Black line above the curves indicates the slope of the Kolmogorov power-law, $E_{k}\propto k^{-5/3}$.}
\label{fig:2nd_jetB}
\end{figure}

\begin{figure}[H]
\centering
\includegraphics[width=8cm]{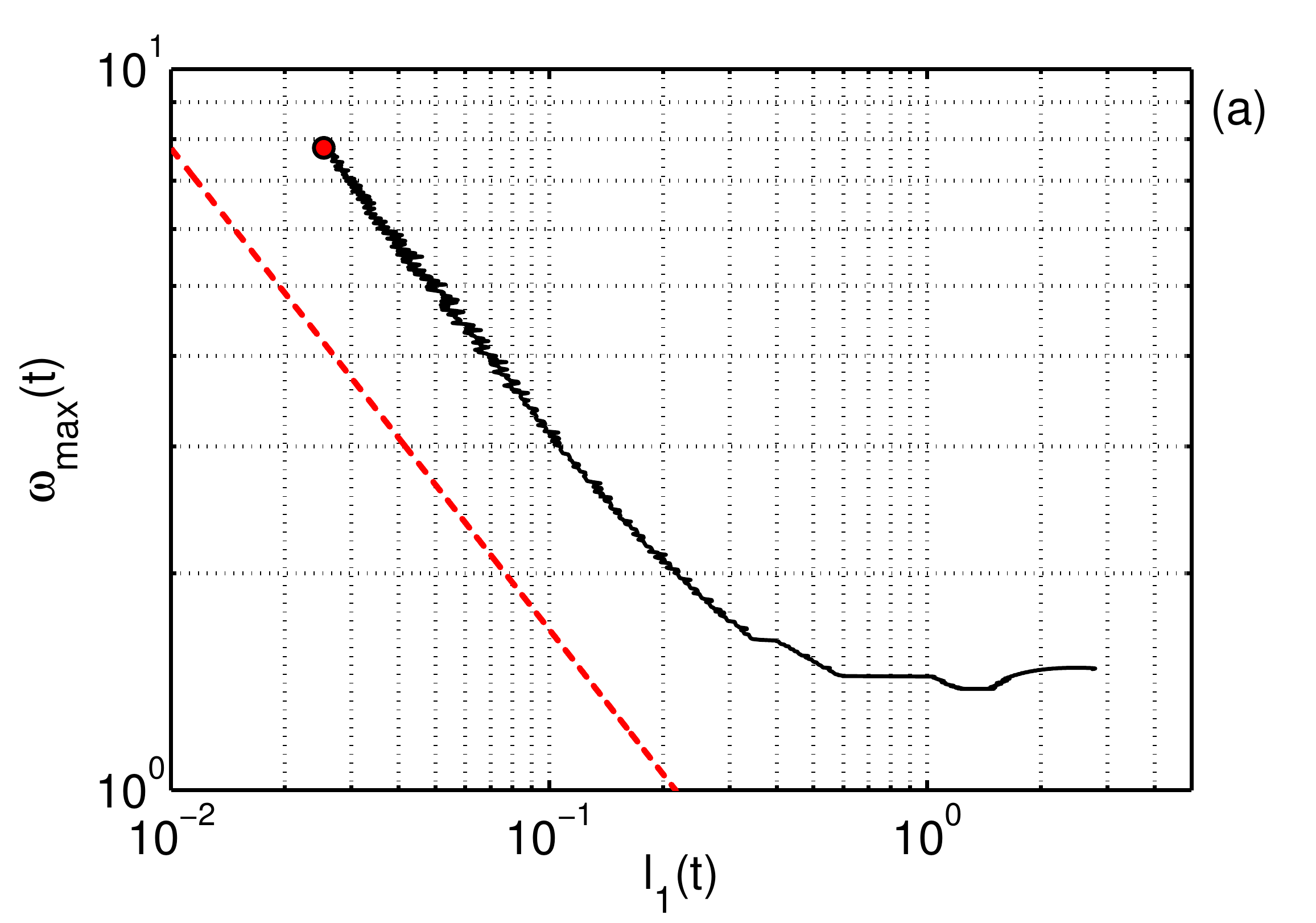}
\includegraphics[width=8cm]{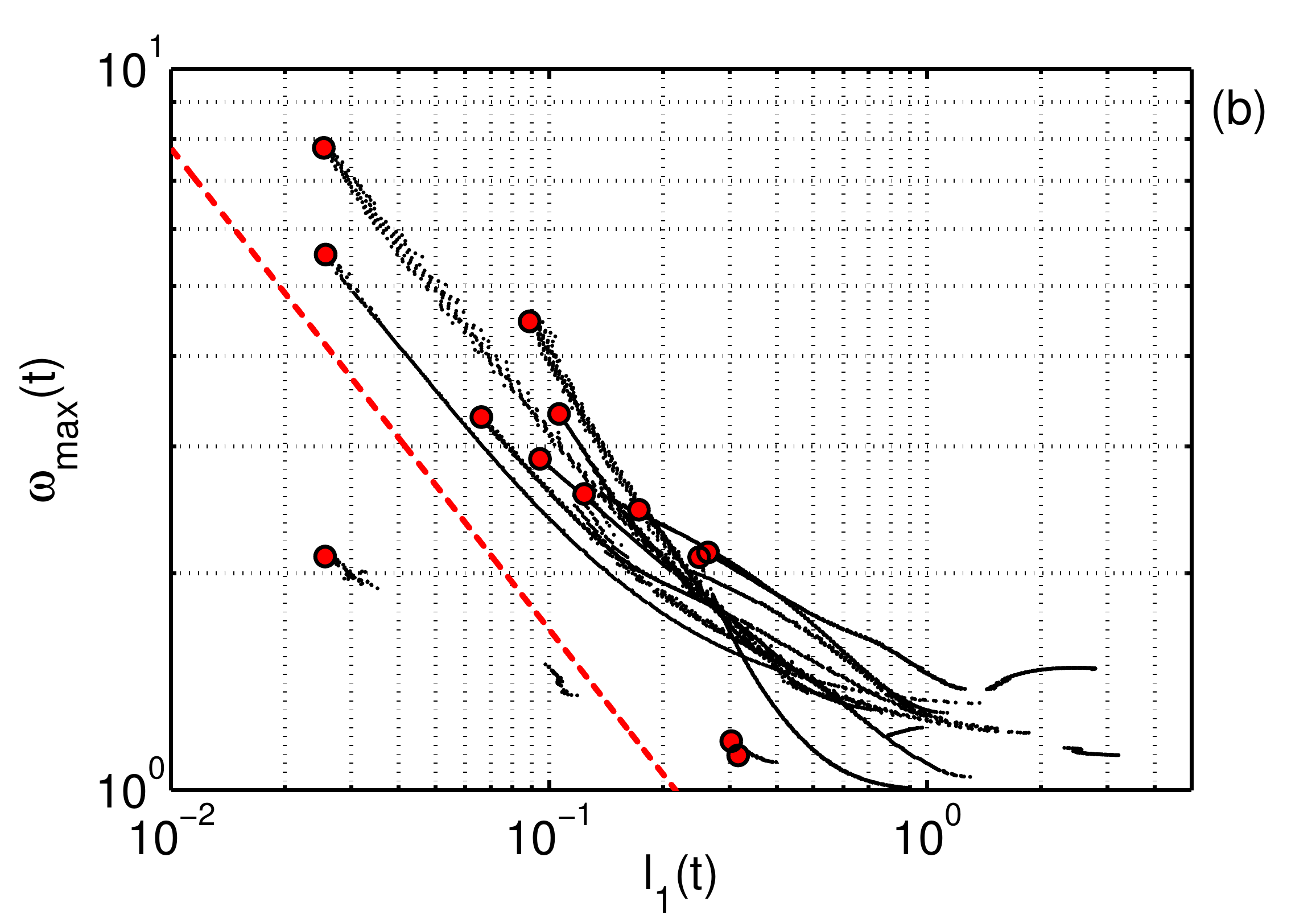}
\caption{{\it (Color on-line)} Relation between the vorticity local maximums $\omega_{\max}(t)$ and the respective characteristic lengths $\ell_{1}(t)$ during the evolution of the pancake structures: (a) for the global maximum, (b) for all local maximums. Red circles mark the local maximums at the final time $t=7.77$. Dashed lines indicate the power-law $\omega_{\max} \propto \ell_1^{-2/3}$ in Eq.~(\ref{singularity_w_to_dk}).}
\label{fig:2nd_W_vs_dk}
\end{figure}

Fig.~\ref{fig:2nd_jet}(a) shows the isosurface of renormalized Fourier components of vorticity $|\tilde{\omegabold}(\mathbf{k})| = 0.2$, see Eq. (\ref{vorticity_renormalized}), where the very thin interior part corresponds to larger vorticity. 
This isosurface is aligned close to the eigenvector $\mathbf{w}_{1}$ computed at the global vorticity maximum. At smaller wavenumbers $k\lesssim 15$, corresponding to the Kolmogorov interval in Fig.~\ref{fig:2nd_evolution_lmB}(a), the isosurface represents a union of several jets,  Fig.~\ref{fig:2nd_jet}(b), that are clearly related to other local vorticity maximums (the figure shows the directions of the first, second and fourth largest local maximums, while the third local maximum yields the direction very close to that for the first one). The small interior part of the isosurface $|\tilde{\omegabold}(\mathbf{k})| = 0.2$ dominates in the energy spectrum, Fig.~\ref{fig:2nd_jetB}(a). The contributions of the jets to the energy spectrum are comparable at sufficiently small wavenumbers from the Kolmogorov region, while at larger wavenumbers (both inside and outside the Kolmogorov region) the contribution from the leading jet becomes dominant, Fig.~\ref{fig:2nd_jetB}(b). 
As shown in Fig.~\ref{fig:2nd_W_vs_dk}, vorticity maximums tend to evolve along the $\omega_{\max}(t)\propto\ell_{1}(t)^{-2/3}$ 
power-law, with rather close values of the constant prefactors before $\ell_{1}^{-2/3}$ for most of them.
Thus, the simulation for the second initial condition leads to the same conclusions on the formation and structure of the 
Kolmogorov turbulent spectrum, as deduced for the first simulation in the main text of the paper.

\end{document}